\documentclass[acmsmall, authorversion]{acmart}

\AtBeginDocument{%
  }

\setcopyright{cc}
\setcctype{by}
\acmJournal{PACMMOD}
\acmYear{2026} \acmVolume{4} \acmNumber{4} \acmArticle{259}
\acmMonth{9} \acmPrice{} \acmDOI{10.1145/3837097}

\usepackage{listings}

\lstdefinestyle{txtprompt}{
    language=,                            %
    basicstyle=\ttfamily\footnotesize,    %
    rulecolor=\color{gray!50},            %
    showspaces=false,                     %
    showstringspaces=false,               %
    numbers=none,                         %
    xleftmargin=0.5em,                      %
    xrightmargin=0cm,                     %
    breaklines=true,                      %
    breakatwhitespace=true,               %
    postbreak=\mbox{\ensuremath{\hookrightarrow}\space}, %
    moredelim=[s][\bfseries]{\{}{\}},  %
    alsoletter={\$},
    morekeywords=[2]{\$DBMS,\$DOCUMENTATION,\$GENERATORS},
    keywordstyle=[2]\bfseries,
    literate={\\n}{{\textcolor{javagreen}{\textbackslash n}}}2
}

\definecolor{javared}{rgb}{0.6,0,0} %
\definecolor{javagreen}{rgb}{0.25,0.5,0.35} %
\definecolor{javapurple}{rgb}{0.5,0,0.35} %
\definecolor{javadocblue}{rgb}{0.25,0.35,0.75} %

\usepackage{multirow}
\usepackage{listings}
\usepackage{color}
\usepackage{svg}
\usepackage{amsmath}
\usepackage[normalem]{ulem}
\usepackage{fontawesome}
\usepackage{csquotes}
\usepackage[ruled,vlined,linesnumbered]{algorithm2e}

\usepackage{tcolorbox}
\usepackage{marginnote}
\usepackage{enumitem}

\newif\ifshowrevisions
\showrevisionsfalse

\newcommand{\revise}[2][]{%
  \ifshowrevisions
    \ifx\relax#1\relax\else%
      \marginnote{%
        \color{black}%
        \fboxsep=3pt%
        \fboxrule=0.5pt%
        \fcolorbox{black!30}{yellow!20}{%
          \begin{minipage}{\linewidth}%
            \centering%
            \small\textup{\texttt{#1}}%
          \end{minipage}%
        }%
      }%
    \fi%
    {\color{blue}#2}%
  \else
    #2%
  \fi
}

\newif\ifshowremoved
\showremovedfalse
\newcommand{\remove}[1]{%
  \ifshowremoved{\color{red!70!black}\sout{#1}}\fi
}

\definecolor{mycolor}{rgb}{0.122, 0.435, 0.698}%
\definecolor{gray1}{gray}{0.3}
\definecolor{darkgreen}{rgb}{0.0, 0.5, 0.0}
\definecolor{darkred}{rgb}{0.82, 0.1, 0.26}
\definecolor{shallowgreen}{RGB}{196, 214, 160}
\definecolor{shallowred}{RGB}{217, 149, 143}
\newcommand{\result}[1]{%
\begin{tcolorbox}[colframe=mycolor,boxrule=0.5pt,arc=4pt,
      left=6pt,right=6pt,top=6pt,bottom=6pt,boxsep=0pt,width=\columnwidth]%
      {#1}
\end{tcolorbox}%
}

\begin{document}
\newcommand{\SQLiteSQLancer}{8.6K}

\newcommand{\ApproachName}{\emph{Approach}}
\newcommand{\ApproachAcronym}{\emph{ShQveL}}
\newcommand{\SystemName}{\emph{ShQveL}}
\newcommand{\ProjectName}{\emph{SQLancer++}}
\newcommand{\DBMSUnderTestNum}{5}
\newcommand{\SQLancer}{\emph{SQLancer}}
\newcommand{\SQLancerSupportDBMSs}{22}
\newcommand{\SQLancerAvgLOC}{3729}
\newcommand{\SupportedDBMSs}{18}
\newcommand{\AvgLOCDBMS}{11.7}

\newcommand{\CPU}{64-core AMD EPYC 7763 CPU at 2.45GHz}
\newcommand{\ProjectLOC}{8.4K}
\newcommand{\SQLancerLOC}{83K}

\newcommand{\SQLiteValidRateWithFB}{98.7\%}
\newcommand{\SQLiteValidRateIncrease}{151.1\%}
\newcommand{\PostgreSQLValideRateIncrease}{52.7\%}
\newcommand{\PostgreSQLValidRateWithFB}{53.6\%}

\newcommand{\LogicBugs}{39}
\newcommand{\LogicBugCockroachDB}{1}
\newcommand{\LogicBugCrateDB}{20}
\newcommand{\LogicBugDuckDB}{13}
\newcommand{\LogicBugMonetDB}{2}
\newcommand{\LogicBugTiDB}{3}
\newcommand{\OtherBugs}{16}

\newcommand{\OverallBugs}{55}
\newcommand{\FixedBugs}{50}
\newcommand{\ConfirmedBugs}{5}
\newcommand{\AvgCrateDBBugswithFB}{26.4}
\newcommand{\AvgCrateDBBugswoFB}{21.2}
\newcommand{\AvgCrateDBUniqueBugswithFB}{8.4}
\newcommand{\AvgCrateDBUniqueBugswoFB}{7.2}

\newcommand{\DuckDBCoverageIncreaseSQLancer}{45\%}
\newcommand{\DuckDBCoverageIncreaseSQLancerpp}{30\%}
\newcommand{\SQLiteValidityDrop}{65.1\%}
\newcommand{\DuckDBValidityDrop}{34.1\%}
\newcommand{\PostgreSQLValidityDrop}{31.8\%}

\title{Automated Database Testing via LLM-Synthesized SQL Features}

\author{Suyang Zhong}\email{suyang@u.nus.edu}
\orcid{0009-0003-0341-7362}
\affiliation{
    \institution{National University of Singapore}
    \city{Singapore}
    \country{Singapore}   
}

\author{Manuel Rigger}\email{rigger@nus.edu.sg}
\orcid{0000-0001-8303-2099}
\affiliation{
    \institution{National University of Singapore}
    \city{Singapore}
    \country{Singapore}   
}

\begin{abstract}
Database Management Systems (DBMSs) have been tested by various automated testing approaches. Many of them generate pairs of equivalent queries to identify bugs that cause DBMSs to compute incorrect results, and have found hundreds of bugs in mature, widely used DBMSs. Most such approaches are based on manually written SQL generators; however, their bug-finding capabilities remain constrained by the limited set of SQL features supported by the generators. In this work, we propose ShQveL, an approach that augments existing SQL test-case generators by leveraging Large Language Models (LLMs) to synthesize SQL fragments. Our key idea is to systematically incorporate SQL features gained through automated interactions with LLMs into the SQL generators, increasing the features covered while efficiently generating test cases. Specifically, ShQveL uses SQL sketches---SQL statements with incomplete code segments that LLMs fill---to integrate LLM-generated content into the generator. We evaluated ShQveL on 5 DBMSs and discovered 55 unique and previously unknown bugs, 50 of which were promptly fixed after our reports.

\end{abstract}


\renewcommand{\shortauthors}{Suyang Zhong and Manuel Rigger}

\received{17 January 2026}
\received[revised]{19 May 2026}
\received[accepted]{12 June 2026}

\maketitle

\section{Introduction}
Database Management Systems (DBMSs) are large, complex, and fundamental software systems. 
Unsurprisingly, they are prone to bugs.
Various approaches have been proposed to detect~\emph{logic bugs} in them using automated testing~\cite{slutz1998massive, rigger2020finding,rigger2020detecting,rigger2020testing, hao2023pinolo, tang2023detecting}. 
They primarily tackle the so-called \emph{test-oracle problem} by validating whether a DBMS operates as expected, for example, by deriving an equivalent query from a given input query to check whether the DBMS produces consistent results~\cite{zhang2025constant, rigger2020finding, rigger2020detecting, ba2024keep}.
Such test oracles have successfully identified hundreds of bugs in popular traditional DBMSs like SQLite and MySQL, as well as emerging DBMSs like DuckDB and TiDB.
Most automated DBMS testing tools automatically derive SQL test cases. They can be broadly classified into generation-based ones, which generate test cases from scratch, based on rule-based generators, while mutation-based ones mutate existing statements.

More than a thousand DBMSs exist,\footnote{https://dbdb.io/} and automatically applying existing testing tools to all of them is challenging.
The key challenge is that DBMSs' dialects differ in both syntax and semantics, including operations and data types, which existing automated testing approaches utilize in their tests. 
To test dialect-specific features, both mutators and generators would ideally generate or mutate instances of these features.
Mutation-based test case generation~\cite{liang2022detecting, jiang2024detecting} depends on high-quality seed inputs, which can be extracted from test suites.
However, tests are often inapplicable to DBMSs other than those for which they were written due to significant dialect differences~\cite{Zhong2025understanding}.
Besides, mutation-based testing tools are difficult to employ to find logic bugs since many test oracles impose strict input constraints.
Conversely, generation-based methods~\cite{rigger2020detecting, rigger2020finding, rigger2020testing} require significant human effort to create DBMS-specific generators.

\ProjectName{} represents an initial attempt to address these challenges~\cite{zhong2025scaling}.
It consists of an SQL generator that infers which features are supported by a given DBMS. It can be used on different DBMSs regardless of SQL dialects without modifications to its source code.
For a specific feature (\emph{e.g.}, a function or an operator), \ProjectName{} executes a number of test cases and then infers the probability of it being supported based on the execution status of the statements containing this feature.
However, \emph{SQLancer++}'s ability to find bugs is constrained by the initially supported set of manually implemented SQL features, potentially missing unique or newly released features of a DBMS.
For example, it supports only 3 data types and 58 functions, and it would be difficult to maintain a generator that supports most features of most DBMSs.
Listing~\ref{listing:example-crate} demonstrates one bug in CrateDB that \ProjectName{} failed to find since it lacks support for the nested data type \texttt{ARRAY(STRING)} and the incorrectly implemented function \texttt{ARRAY\_POSITION}.
Supporting new DBMS-specific features is a labor-intensive process that requires expertise in both the DBMS and the test case generator. 
Thus, an automatic method to integrate these features into the generator is essential.

\begin{lstlisting}[language=SQL,float=tb,  caption={An example bug in CrateDB caused by scalar function \texttt{ARRAY\_POSITION}. The query is expected to return an empty result; however, one redundant row is returned.}, label={listing:example-crate}, escapechar=@, 
showspaces=false,
numbers=left,
xleftmargin=1em,
xrightmargin=-0.5cm,
numberstyle=\scriptsize,
commentstyle=\color{gray},
keywordstyle=\color{black}\bfseries,
breaklines=true,
basicstyle=\ttfamily\footnotesize,
belowskip=-1em,
breakatwhitespace]
CREATE TABLE t0(c0 INT, c1 ARRAY(STRING));
INSERT INTO t0(c0, c1) VALUES (1, ['a', 'b']);

SELECT c0 FROM t0 
WHERE (c0 != ARRAY_POSITION(t0.c1, 'c', 1));
-- {1} @\faBug@ -- {} @\faCheck@
\end{lstlisting}

Recent progress on \emph{Large Language Models (LLMs)} suggests the possibility of using them as DBMS test-case generators.
First, prior research has applied LLMs to data management, such as in the context of text-to-SQL~\cite{li2023can, gao2024text} and query rewriting~\cite{li2024llmr2, liu2024query}.
LLMs have also been applied to testing in a variety of domains, including general-purpose testing~\cite{xia2024fuzz4all}, compiler testing~\cite{yang2024whitefox}, and testing of deep learning libraries~\cite{deng2023large}.
However, several challenges persist when applying LLMs to test DBMSs.
First, the throughput of LLM-based generation is significantly lower than that of traditional query generators.
For instance, LLM-R\textsuperscript{2}~\cite{li2024llmr2} leverages LLMs to rewrite queries, resulting in an average latency exceeding one second per query on each of its benchmarks, while tools like \SQLancer{} can generate thousands of queries per second.
Second, the cost of employing LLMs for testing is high. 
Utilizing an LLM typically requires access to a powerful GPU server or incurs significant expenses through API services (\emph{e.g.}, GPT-4o costs 15 US dollars for every 1 million tokens).
As a result, integrating LLMs into the CI/CD pipelines for DBMS development is neither straightforward nor cost-effective.
Third, current LLMs suffer from hallucination~\cite{huang2025survey, zhang2023siren}, producing unreliable outputs.
In summary, the inherent inefficiency, expense, and unreliability of current LLMs pose significant challenges to their application in the testing domain.

In this paper, we propose \emph{ShQveL}, a technique that enables existing SQL generators to incorporate features via LLMs.
Our core insight to achieve \emph{efficient} SQL generators that can generate DBMS-specific SQL features is to leverage LLMs by permanently integrating the LLM-generated content after automatically validating it.
After an initial learning phase, we can disable LLM interactions and thus achieve an efficiency comparable to manually written generators.
A key challenge is to design the approach such that LLMs can be used to determine DBMS-specific features, and subsequently validate and integrate this knowledge into the SQL generators.
To bridge the gap between the LLM-generated content and generator source code, we propose the concept of \emph{SQL sketching}, loosely inspired by the concept of sketch-based program synthesis~\cite{solar2008program, wang2019synthesizing}.
An SQL sketch is a template of SQL statements generated by the original SQL generator with incomplete segments as placeholders, or \emph{holes}.
During learning, an LLM fills the holes in the sketch based on its pre-trained knowledge and DBMS documentation through in-context learning.
Subsequently, the LLM-generated fragments are integrated into the generator.
In order to impose constraints on the LLM, we prompt it by providing a SQL sketch that implicitly imposes constraints on feature usage (\emph{e.g.}, a SQL sketch \texttt{SELECT 1 ?? 1} indicates that the operator should have two integer operands).
After receiving the LLM's response, our approach validates the resulting filled sketch and subsequently uses the validated fragments during testing.

We implemented \SystemName{} based on SQLancer++ and evaluated it on CrateDB, CockroachDB, DuckDB, MonetDB, and TiDB.
We found and reported \OverallBugs{} unique, previously unknown bugs, showing the effectiveness of our approach. \FixedBugs{} bugs have been fixed, demonstrating that the developers considered the bugs important.
Our goal was not to outperform the manually written generators (\emph{e.g.}, \SQLancer{}), but to find bugs in DBMSs of different dialects without investing any implementation effort.
Despite this, \SystemName{} achieved comparable code coverage on PostgreSQL and SQLite, and, on DuckDB, observed increases of \DuckDBCoverageIncreaseSQLancer{} compared with SQLancer++ and \DuckDBCoverageIncreaseSQLancerpp{} compared with SQLancer.
ShQveL can learn features efficiently and economically, as it learned around 400 fragments for each DBMS in six hours with a cost of less than 1 USD per DBMS.

We believe that our approach is both practical and scalable. 
The learned fragments can persist across executions and can be reused even across different systems.
Once a sufficient number of features has been learned, the learning process can be stopped, leading to more efficient test-case generation, especially in resource-limited environments (\emph{e.g.}, CI runs).
Additionally, our \ProjectName{}-based feedback mechanism validates LLM-generated fragments without human supervision.
In general, this approach makes the testing process more efficient, economical, and manageable.

In summary, we make the following contributions:
\begin{itemize}
    \item We propose \emph{SQL sketching} as a general notion for integrating LLM-generated content with generation-based DBMS testing.
    \item We propose a testing approach, \emph{ShQveL}, which leverages LLMs to persist and integrate DBMS-specific features into the generator.
    \item We implemented and evaluated the approach, which has found \OverallBugs{} unique, previously unknown bugs in widely used DBMSs.
\end{itemize}

\begin{table}[tb]
    \centering
\footnotesize
\caption{Comparison of an LLM-based generator and a manually written generator on DuckDB for 6 hours.}\label{tab:duckdb-time}
\begin{tabular}{ llrrrrr }
\toprule
\multirow{2}{*}{Method} & \multirow{2}{*}{Device}&Unique Bugs   & Branch  &  Generated  &  Validity \\
    &                & Found & Cov. &Stmts.  & Rate  \\
\midrule
Fuzz4All       & GPU & 0 & 24.5\%  &  183K &  70.9\% \\
SQLancer++     & CPU & 2 & 21.3\% & 41,976K & 93.1\% \\
\bottomrule
\end{tabular}

\end{table}

\section{Background and Motivation Study}
To understand the rationale behind our approach, we first introduce the preliminaries for generating SQL test cases in a dialect-agnostic manner (see Section~\ref{sec:sqlancer++}).
Subsequently, we demonstrate the challenges of using LLMs in DBMS testing and the limitations of two naive approaches: (1) generating test cases directly with the LLM (see Section~\ref{sec:LLM-testcase}) and (2) generating test generators with the LLM (see Section~\ref{sec:LLM-generator}).
Given the LLMs' proficiency in SQL tasks like text-to-SQL, a straightforward approach is to have them generate SQL statements for fuzzing, as described in a recent blog post.\footnote{https://celerdata.com/blog/chatgpt-is-now-finding-bugs-in-databases}
For example, we can prompt the LLM to generate SQL statements based on the corresponding documentation and execute the generated statements on the target systems.
LLMs have also achieved high performance on code generation~\cite{jimenez2024swebench, GitHubCopilot}, suggesting that they can be used to synthesize generators, a possibility that we thus explored.
The throughput of executing synthesized generators that themselves eschew interactions with LLMs is expected to be higher, but any error in an LLM's output would render the whole generator unusable.
We chose DuckDB v0.7.1, a historical version of the widely used embedded relational DBMS, as our target~\cite{rigger2020finding, fu2024sedar}.

\subsection{Dialect-agnostic SQL Generation}\label{sec:sqlancer++}
\ProjectName{} is an automated platform for testing DBMSs to find logic bugs, aiming to test a wide range of DBMSs with different SQL dialects.
Its core component is an adaptive SQL generator that, during execution, dynamically infers which of a set of predefined SQL features---elements or properties in the query language that can be specified at different levels of granularity---are supported by a given DBMS.
Specifically, a feature might be a keyword, such as \texttt{DISTINCT}, or an entire SQL statement, like \texttt{VACUUM}.
It can also refer to operators or functions, for example, the null-safe operator \texttt{<=>} in MySQL.
To infer whether this feature is supported by the system under test, during execution, SQLancer++ generates a sufficient number of SQL statements containing the feature.
Subsequently, SQLancer++ calculates the estimated probability of the features being supported based on whether the DBMS executed a predefined percentage of instances of the features successfully in previous executions.
SQLancer++ supports only a limited set of SQL features across DBMSs, as it relies on a manually selected subset of basic, widely shared features across dialects.
It is feasible to add more features for one specific DBMS; however, it requires a high implementation effort, and it would be infeasible to add and maintain dialect-specific features of many DBMSs, thus motivating the subsequent exploration.

\subsection{LLM-based Test Generation}\label{sec:LLM-testcase}
\paragraph{Methodology}
We first demonstrate the potential and challenges of using LLMs to directly generate test cases by applying Fuzz4All~\cite{xia2024fuzz4all} to test DuckDB.\sloppy{}
Fuzz4All is a universal state-of-the-art fuzzer that leverages LLMs (\emph{e.g.}, StarCoder~\cite{li2023starcoder}) to target a variety of input languages, including C/C++ and Java, although it was not specifically evaluated on SQL.
We executed both Fuzz4All and SQLancer++ continuously for 6 hours on a GPU server, comparing their test efficiency and test case validity rates.
For Fuzz4All, we configured the system with a summary of the DuckDB documentation while keeping all other settings at their defaults (\emph{e.g.}, GPT-4 as distillation LLM and StarCoder as generation LLM). 
We executed both systems in a single-threaded manner for comparability.

\paragraph{Results}
Table~\ref{tab:duckdb-time} summarizes the results of our experiments.
First, SQLancer++ generated 229$\times$ more SQL statements, as invoking the LLM for every SQL statement is inefficient.
Second, we observed a higher validity rate for SQLancer++, although it also fails to achieve a 100\% validity rate, because certain features can trigger expected errors due to semantic constraints (\emph{e.g.}, adding two very large integers may overflow and result in an error).
Third, SQLancer++ reported 156 bug-inducing test cases with 2 unique bugs compared to none discovered by the LLM-based approach.
However, the LLM-based method achieved higher branch coverage; one plausible explanation is that it can generate DuckDB-specific SQL features, while SQLancer++ was designed and implemented to support mostly common ones.
While exercising code is a necessary condition for finding a bug, it is not sufficient; in practice, only specific test cases might trigger it, which is why high throughput is also necessary~\cite{bohme2020boosting}.

\result{
LLM-generated SQL test cases incorporating DBMS-specific features can achieve higher coverage, while manually written generators with higher throughput can detect more bugs.
}

\subsection{LLM-based Generator Generation}\label{sec:LLM-generator}
\paragraph{Methodology}
We employed LLMs to synthesize a statement generator for DuckDB to explore the potential of efficient test case generation.
We used ChatGPT o1---one of the most advanced models for code at the time of writing---to synthesize a program capable of generating random \texttt{CREATE TABLE} statements with the following prompt:
\emph{``Act as an advanced Python and database developer. Please generate a Python script that, upon execution, randomly creates a DuckDB CREATE TABLE statement.''}
Additionally, we provided relevant documentation to guide the model in generating DBMS-specific content.

\begin{lstlisting}[language=Python,float=tb,  caption={A random DuckDB \texttt{CREATE TABLE} generator synthesized by ChatGPT o1. We shortened variable names and omitted helper functions for simplicity.}, label={listing:gpt-generator}, escapechar=@,
showspaces=false,
numbers=left,
xleftmargin=1em,
xrightmargin=-0.5cm,
numberstyle=\scriptsize,
commentstyle=\color{javagreen},
keywordstyle=\color{black}\bfseries,
breaklines=true,
basicstyle=\ttfamily\footnotesize\linespread{0.8},
breakatwhitespace]
def rand_constr(col_name, existing_cols):
    constrs = []
    col_constrs = [
        f"UNIQUE",
        # Issue 1: Value is fixed
        f"CHECK ({col_name} > 0)",
        f"CHECK ({col_name} <> '')",
    ]
    use_col_constraints = random.choice([True, False])
    if use_col_constraints:
        constrs.append(random.choice(col_constrs))
    if existing_cols and random.choice([True, False]):
        referenced_col = random.choice(existing_cols)
        # Issue 2: SQL syntax is incorrect
        # Issue 3: Table name is hardcoded
        constrs.append(f"FOREIGN KEY ({col_name}) REFERENCES some_other_table({referenced_col})")
    return " ".join(constrs)

def generate_create_table_statement():
    # Generate columns_sql with rand_constr
    # ...
    # Construct the SQL statement and return it
    return f"CREATE TABLE {table_name} (\n    {columns_sql}\n);"
\end{lstlisting}

\remove{
\begin{lstlisting}[language=SQL,float=tb,  caption={A statement generated by the synthesized generator.}, label={listing:example-crtab}, escapechar=@,
showspaces=false,
numbers=left,
xleftmargin=1em,
xrightmargin=-0.5cm,
numberstyle=\scriptsize,
commentstyle=\color{gray},
keywordstyle=\color{black}\bfseries,
breaklines=true,
basicstyle=\ttfamily\footnotesize,
breakatwhitespace]
CREATE TABLE TABLE_QBLVC (
    COL_EQVWT BIGINT CHECK (COL_EQVWT <> ''),
    COL_KKXBL DATE FOREIGN KEY (COL_KKXBL) REFERENCES SOME_OTHER_TABLE(COL_EQVWT)
);
\end{lstlisting}}

\paragraph{Results}
Listing~\ref{listing:gpt-generator} shows the code that ChatGPT synthesized\remove{, and Listing~\ref{listing:example-crtab} shows an example of its output}.
While ChatGPT’s output is almost correct, we observed that hallucinations in the LLM’s output can cause bugs.
For example, lines 13–17 of Listing~\ref{listing:gpt-generator} incorrectly apply a \texttt{FOREIGN KEY} constraint as a column constraint in DuckDB, which may result in a syntax error when executing the generated SQL statement\remove{ (see Listing~\ref{listing:example-crtab})}.
Although we supplied DuckDB's documentation, ChatGPT still produced incorrect code---likely due to difficulties in parsing highly structured languages~\cite{wang2023grammar}. 
Such errors could potentially render the whole testing tool unusable unless they are manually fixed.

An additional challenge is how to assemble multiple such generators to form a complete automated testing tool, as well as how to address concerns such as schema management, value diversity, and statement dependencies.
First, LLMs struggle to maintain consistent schema references. For example, LLMs might not reference tables created by the \texttt{CREATE TABLE} statements.
In \revise[R2.M1]{line~16} of Listing~\ref{listing:gpt-generator}, the generator references a non-existent table, \texttt{some\_other\_table}.
Second, generators synthesized using LLMs often explore a limited part of the search space; \revise{line~6} of Listing~\ref{listing:gpt-generator} compares a column’s value to 0, while it may overlook boundary values (\emph{e.g.}, the maximum integer value) that may trigger bugs.
Third, statements must be executed in the correct order; for example, \texttt{CREATE TABLE} should precede \texttt{INSERT}.
While advanced approaches in LLM agents for general coding tasks~\cite{zhang2024autocoderover, yang2024swe} might address some of these issues, additional manual effort will likely still be required, and subtle errors could affect the effectiveness and efficiency.

\result{
LLMs struggle to synthesize SQL generators correctly, as they face internal hallucination issues and other challenges---schema handling, generating diverse and valid constants, and statement scheduling---which together lead to suboptimal accuracy and reliability in the generated queries.
}

\section{\SystemName{}}
We propose \SystemName{}, an approach to automatically test DBMSs, supporting a multitude of SQL dialects through LLM guidance.
The core idea behind our approach is to use LLMs to extract knowledge of different dialects' SQL features and corresponding DBMS documentation to then integrate this knowledge into SQL generators automatically. 
Consequently, unlike existing solutions like \SQLancer{} or \ProjectName{}, \SystemName{} can potentially cover each DBMS's SQL features even without requiring manual implementation effort.
To avoid the high cost and low throughput of directly instructing the LLM to generate SQL tests (see Section~\ref{sec:LLM-testcase}), we persist knowledge gained through LLM interactions; after a user has decided that a sufficient number of features has been learned, no further LLM invocations are required.  %
Due to hallucinations and other limitations (\emph{e.g.}, limited context length) of current LLMs, directly synthesizing SQL generators would result in errors, requiring manual intervention (see Section~\ref{sec:LLM-generator}).
Rather, we propose \emph{SQL sketches} for generator augmentation, in which a manually written \emph{base generator} generates basic, common SQL statements, some of which contain \emph{holes} to be filled by the LLM with code fragments of DBMS-specific features.
The fragments synthesized are validated by a self-validation mechanism to eliminate invalid fragments caused by LLM hallucinations.
To address the challenge that LLM-synthesized fragments are unaware of valid tables, columns, or types from the current database schema, and may produce deterministic rather than diverse outputs, we provide the LLM with schema information and random literal generators as part of the prompt.

\begin{figure*}[tb]
    \centering
    \includegraphics[width=0.95\linewidth]{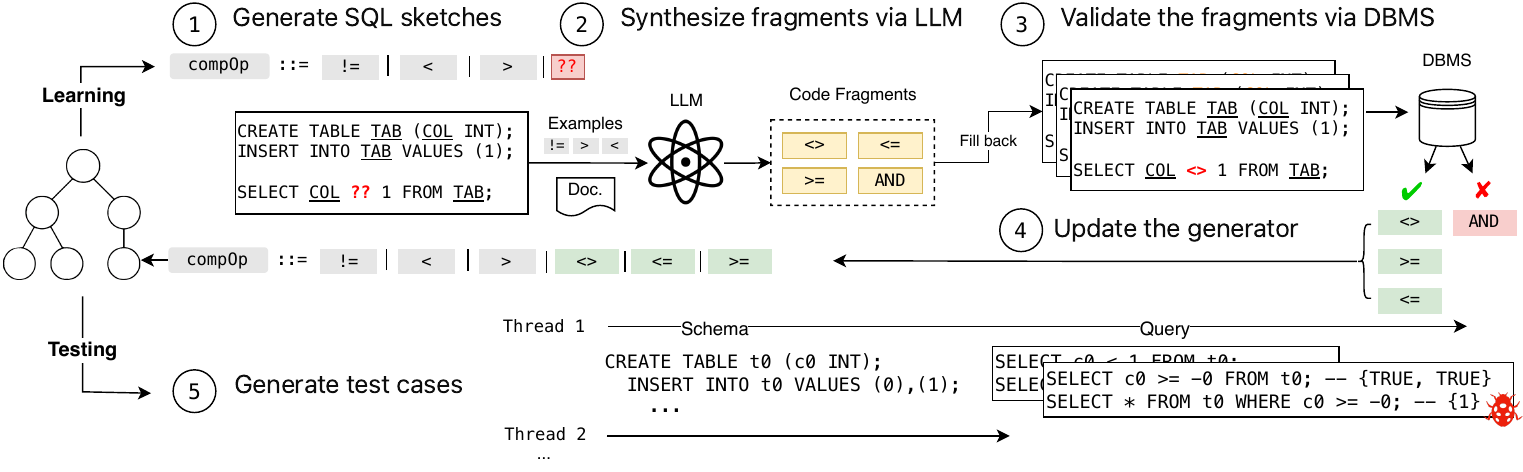}
    \caption{Overview of \ApproachAcronym{}}
    \label{fig:mta}
\end{figure*}
\paragraph{System overview}
Figure~\ref{fig:mta} shows an overview of \emph{ShQveL}.
During the execution of \SystemName{}, the base generator is either instructed to initiate the learning phase, aiming to learn new features, in which case \SystemName{} generates a SQL sketch (see step \textcircled{1}), or \SystemName{} generates test cases aiming to find bugs in the DBMS (see step \textcircled{5}).
In the learning phase, \SystemName{} invokes an LLM to fill the SQL sketch (see step \textcircled{2}).
One SQL sketch can contain multiple incomplete segments.
Each incomplete segment \texttt{??}---referred to as a \emph{hole}---can be filled with DBMS-specific feature fragments (\emph{e.g.}, SQL keywords, constants, or expressions).
We refer to each LLM-synthesized string that can be substituted into a hole as a \emph{fragment}, and a sketch whose holes are all filled is a \emph{filled sketch}.
\SystemName{} then invokes an LLM to fill the SQL sketch, using the target DBMS documentation and few-shot examples as guidance.
The LLM fills the holes with fragments, and the resulting filled sketch is then executed on the system for validation (see step \textcircled{3}).
If execution is successful, we mark the fragments as \emph{valid}; otherwise, we mark them as \emph{invalid}.
In step \textcircled{4}, \SystemName{} updates the generator using the validated fragments.
Based on the updated generator, \SystemName{} keeps generating SQL statements to test the target system (see step \textcircled{5}). 

\begin{table*}[tb]
    \centering
\small
\caption{SQL sketch design for \SystemName{}}\label{table:sketches}
\resizebox{\columnwidth}{!}{%
\begin{tabular}{ llll }
\toprule
Feature & Generation Rule & Example Sketch & Example Filling \\
\midrule
Statement& Insert \textbf{??} as a whole statement & \texttt{CREATE TABLE \underline{TAB} (\underline{COL} INT);} &  \texttt{CREATE TABLE \underline{TAB} (\underline{COL} INT);}       \\
      & & \texttt{\textbf{\textbf{??}}}; & \texttt{\textcolor{red}{\textbf{ANALYZE}}}; \\
Clause & Insert \textbf{??} into statements  &  \texttt{CREATE TABLE \textbf{\textbf{??}} \underline{TAB}} & \texttt{CREATE TABLE \textcolor{red}{\textbf{IF NOT EXISTS}} \underline{TAB}}         \\
&&\texttt{(\underline{COL} INT \textbf{\textbf{??}}, \textbf{\textbf{??}})\textbf{\textbf{??}};} & \texttt{    (\underline{COL} INT \textcolor{red}{\textbf{NOT NULL}}, \textcolor{red}{\textbf{PRIMARY KEY (\underline{COL})}});}\\
Data type & Insert \textbf{??} as data type name and value  & \texttt{CREATE TABLE \underline{TAB} (\underline{COL} \textbf{\textbf{??}});}  & \texttt{CREATE TABLE \underline{TAB} (\underline{COL} \textcolor{red}{\textbf{ARRAY}});}       \\
 && \texttt{INSERT INTO \underline{TAB} VALUES (\textbf{\textbf{??}});}  &   \texttt{INSERT INTO \underline{TAB} VALUES (\textcolor{red}{\textbf{[1, <RANDOM\_INT>]}});}    \\ 
Expression  & Insert \textbf{??} as expression node & \texttt{CREATE TABLE \underline{TAB} (\underline{COL} ARRAY);}   &  \texttt{CREATE TABLE \underline{TAB} (\underline{COL} ARRAY);}          \\
 &&\texttt{SELECT \textbf{\textbf{??}}(\underline{COL}) FROM \underline{TAB};} & \texttt{SELECT \textcolor{red}{\textbf{ARRAY\_SUM}}(\underline{COL}) FROM \underline{TAB};}\\
\bottomrule
\end{tabular}
}

\end{table*}

\subsection{Sketch Generation}
In step \textcircled{1}, \SystemName{} generates multiple SQL sketches for learning and integrating specific SQL features into the generator.
\revise[R2.O2]{%
The underlying base generator generates concrete SQL statements by sequentially appending string fragments---SQL keywords, schema references, operators, and randomly generated constants.
For example, a typical statement is assembled by concatenating the strings \texttt{SELECT}, \texttt{col0}, \texttt{>}, \texttt{42}, \texttt{FROM} and \texttt{t0}.%
}
To create a SQL sketch, \SystemName{} inserts a hole at developer-specified locations in the generated SQL statements for the feature to be learned.
For example, as shown in Figure~\ref{fig:mta}, the generator might generate a sketch that allows the LLM to specify a binary operator by outputting the placeholder \texttt{??} instead of a concrete operator.
The generator might thus generate an incomplete \texttt{SELECT} statement (e.g., \texttt{SELECT COL ?? 1 FROM TAB}), which the LLM can subsequently fill by synthesizing one of many plausible operators.

\paragraph{Sketch design}~\label{sec:sketch-design}
Table~\ref{table:sketches} shows the example sketches and their corresponding fillings at different levels.
One or more holes (denoted by ??) can exist in one sketch. %
\revise[R3.Q3]{%
The four hole types---\emph{statement}, \emph{clause}, \emph{expression}, and \emph{data type}---mirror the SQL feature taxonomy that \ProjectName{} adopts to enumerate dialect-specific SQL features~\cite{zhong2025scaling}.
Automatically inferring new hole locations from feedback is left as future work (Section~\ref{sec:discussion}).%
}
For clause-level features, the developer can replace any segment of an SQL statement with a hole or insert a hole into the statement; the LLM then fills the hole, and \SystemName{} integrates the fragment into the generator.
For statement-level features, the entire SQL statement serves as a hole for the LLM to fill and incorporate, and for other feature levels, the developer specifies hole placement \revise[R2.O2]{by annotating specific sites in the generator source (Section~\ref{sec:implementation})}.

\paragraph{Context statements}
To help the LLM synthesize fragments that depend on database state, each SQL sketch includes complete SQL statements that provide context in addition to statements with holes.
These context statements convey schema information to the LLM. For example, a \texttt{SELECT} statement that queries an existing table typically requires preceding \texttt{CREATE TABLE} statements.
In Figure~\ref{fig:mta}, where \SystemName{} is learning the comparison operator feature, the \texttt{SELECT} statement with a hole appears below the \texttt{CREATE TABLE} and \texttt{INSERT} statements, which establish the database state as table \texttt{TAB} and insert sample data into column \texttt{COL}.
Conversely, when LLM-synthesized fragments contain \texttt{COL} or \texttt{TAB}, the generator can identify them as database objects. Consequently, the generator can abstract concrete column references in a fragment so that, during testing, it instead references a column available in the current schema.
As a result, fewer potentially ambiguous natural language instructions are needed in the prompt design, and the filled sketch can be executed to validate that the filled fragments are correct.

\paragraph{Fragment integration}
\revise[R2.O3]{%
Our sketch design enables ShQveL to reuse LLM-learned fragments across feature levels.
After a filled sketch passes validation, \SystemName{} integrates each synthesized fragment by attaching it to the hole that generated it.
Each hole samples from a validated fragment pool; data-type holes store paired type-value templates, while statement, clause, and expression
holes store string fragments.
For example, after learning \texttt{ARRAY} and corresponding insertable values from the data-type sketch in Table~\ref{table:sketches}, row~3, the generator can use \texttt{ARRAY} as a column type in the context statements for expression-level sketches (row~4).
The LLM can then synthesize fragments such as \texttt{ARRAY\_SUM(\underline{COL})} that operate on the learned type.
}

\subsection{Prompt Design and Fragment Synthesis}
\label{sec:fragmentsyn}
In step \textcircled{2}, \ApproachAcronym{} prompts the LLM to fill the holes in the sketches using in-context learning~\cite{brown2020language}.
Two challenges for designing prompts limit this synthesis.
First, the LLM's training data might lack less common SQL dialects or recent SQL features.
Second, LLMs tend to generate outputs based on common patterns seen in their training data.
As further detailed in this section,  \ApproachAcronym{} uses \emph{Retrieval-Augmented Generation (RAG)} techniques to address the first challenge, and incorporates random literal generators to address the second challenge.
Specifically, \ApproachAcronym{} prompts the synthesis LLM with up-to-date DBMS information through the documentation of DBMSs based on RAG techniques.
Additionally, \ApproachAcronym{} exposes an interface to the LLM that allows it to generate random literals to increase the exploration of interesting and diverse test cases.

\paragraph{Prompt design}
We subsequently elaborate on the prompt that \SystemName{} uses to interact with the LLMs (see Listing~\ref{listing:example-prompt}).
Each synthesis prompt asks the LLM to fill the numbered placeholders \texttt{\{0\}}, \texttt{\{1\}}, \dots{} in the generator-produced SQL sketch.
At a high level, the prompt combines lightweight natural-language requirements (\emph{e.g.}, dialect constraints and determinism) with structured context, including the target DBMS, an up-to-date documentation summary via RAG, and the SQL sketch with its context statements.
We implement this design by including a small few-shot set of previously validated fragments and instructing the LLM to output multiple candidates in a simple CSV-like format (\emph{one fragment per line}: \texttt{hole\_id;fragment})\remove{, as illustrated in Listing~\ref{listing:example-answer}}.
This format enables deterministic parsing and direct substitution into sketches for validation (Section~\ref{sec:sketch-validation}).
We also explicitly instruct the LLM to avoid non-deterministic functions; any remaining non-deterministic fragments are handled by the keyword filter and false-alarm analysis described in Section~\ref{sec:fp}.

\begin{lstlisting}[float=tb,  caption={An example prompt for fragment synthesis. \texttt{\textbf{\$DBMS}} represents the target DBMS name, \texttt{\textbf{\$DOCUMENTATION}} represents the summarized documentation, and \texttt{\textbf{\$GENERATORS}} represents the available literal generators (Table~\ref{tab:generators}).}, label={listing:example-prompt}, style=txtprompt]
You are an expert in SQL dialects. Given the following SQL sketch in $DBMS
 
CREATE TABLE TAB (COL ARRAY(STRING));
INSERT INTO TAB VALUES (['a', 'b']);
INSERT INTO TAB VALUES ({0});
INSERT INTO TAB VALUES ({1}(NULL));
INSERT INTO TAB VALUES ({2}(NULL, NULL));
SELECT {3}(COL) FROM TAB;

Task: For each placeholder ({0}, {1}, ...) in the SQL sketch, generate as many deterministic, rare, and complex concrete alternatives as possible...

Refer to examples from the documentation: $DOCUMENTATION

When helpful, use generator identifiers from $GENERATORS in your fragments.
\end{lstlisting}

\remove{
\begin{lstlisting}[float=tb,  caption={An example answer for Listing~\ref{listing:example-prompt}. The returned fragments often reuse feature names appearing in \texttt{\$DOCUMENTATION} (Listing~\ref{listing:rag-documentation}).}, label={listing:example-answer}, style=txtprompt]
0;['c', 'd']\n0;['a', <RANDOM_STRING>]\n...
1;ARRAY_UNIQUE\n1;COALESCE\n...
2;ARRAY_UNNEST\n;ARRAY_CAT\n...
3;ARRAY_SUM\n2;ARRAY_MAX\n2;ARRAY_POSITION\n...
\end{lstlisting}}

\paragraph{Retrieval-augmented generation}
\ApproachAcronym{} uses a RAG-based approach to incorporate up-to-date information from DBMS documentation into the synthesis prompt (see the \texttt{\$DOCUMENTATION} field in Listing~\ref{listing:example-prompt}).
Although relying solely on the LLM's internal knowledge is possible, its performance can be limited, especially for emerging DBMSs whose specifications would likely be missing in the model's training data (see Section~\ref{sec:coverage}). 
LLMs are pre-trained and updated infrequently; thus, they cannot capture the latest knowledge of the target DBMS. 
Besides, their capacity limits their ability to provide detailed, accurate information for less common DBMSs.

\revise[R3.Q4]{%
To address the above issues, \ApproachAcronym{} uses a three-stage retrieval-augmentation process, as shown in function \texttt{rag} in Algorithm~\ref{alg:fuzzing} (lines~\ref{algoline:ragstart}--\ref{algoline:ragend}) and illustrated here with a running example $D=\texttt{CrateDB}$, $f=\texttt{ARRAY}$, $\ell=$ data type.
First, \SystemName{} concatenates $D$, $f$, and a short descriptor of the feature level into a natural-language query (\emph{e.g.}, \texttt{``CrateDB ARRAY data type overview''}).
Second, \SystemName{} submits the query to a search engine and fetches the top $N$ official DBMS documentation pages for $f$ ($N = 3$ in our evaluation); the same component could alternatively be implemented using embedding-based retrieval by indexing the documentation in a vector database.
Third, \SystemName{} invokes a lightweight LLM (\emph{e.g.}, GPT-4o-mini) to distill the retrieved pages into a compact reference summary (\texttt{\$DOCUMENTATION}) that lists, for every feature the page describes, a triple \texttt{(name, description, concrete SQL example)}, reducing the reference size to fit within LLM context limits.%
}

\remove{
\begin{lstlisting}[float=tb,  caption={A prompt template for summarizing raw documentation content into \texttt{\$DOCUMENTATION} (here, \texttt{feature=ARRAY} and \texttt{DBMS=CrateDB}).}, label={listing:rag-summarize-prompt}, style=txtprompt]
You are given a DBMS documentation page content. What are the {feature}-related built-ins for {DBMS} based on the documentation:

{context}

Instructions:
- Return ONLY names and examples (no explanations).
- List all possible values you can find in the provided context.
- Output format: one item per line as "NAME - EXAMPLE".
\end{lstlisting}}

\remove{
\begin{lstlisting}[float=tb,  caption={An example (shortened) \texttt{\$DOCUMENTATION} snippet produced by the summarization LLM for CrateDB \texttt{ARRAY}.}, label={listing:rag-documentation}, style=txtprompt]
ARRAY_UNIQUE - ARRAY_UNIQUE([1, 1, 2]) = [1, 2]
ARRAY_CAT - ARRAY_CAT([1, 2], [3]) = [1, 2, 3]
ARRAY_POSITION - ARRAY_POSITION([10, 20, 30], 20) = 2
ARRAY_SUM - ARRAY_SUM([1, 2, 3]) = 6
...
\end{lstlisting}}

\begin{table}[tb]
    \centering
    \caption{Examples of the random generators.}
    \begin{tabular}{ll}
    \toprule
    Generator Identifier & Description \\
    \midrule
       \texttt{RANDOM\_INT}  &  A random integer, \emph{e.g.}, 1\\
       \texttt{RANDOM\_VARCHAR}  &  A random string, \emph{e.g.}, abc\\
       \texttt{RANDOM\_DATE} & A random date, \emph{e.g.}, 2024/12/13 \\
       \texttt{RANDOM\_TABLE} & A random table reference. \\
       \texttt{RANDOM\_COLUMN} & A random column reference. \\
    \bottomrule
    \end{tabular}
    \label{tab:generators}
    \vspace{-3mm}
\end{table}

\paragraph{Random generators}
\SystemName{} exposes a function interface to the LLM for generating random literals to explore a larger search space during testing (see \texttt{\$GENERATORS} in Listing~\ref{listing:example-prompt}, instantiated from Table~\ref{tab:generators}).
LLMs are likely to generate answers that are frequently seen in their training data, while in testing, using values that are not frequently seen may trigger bugs.
See Listing~\ref{listing:gpt-generator} Issue 1 as an example. 
LLMs can return concrete fragments (\emph{e.g.}, \texttt{CHECK (col\_name > 0)} for a column constraint clause); however, a bug may be triggered only when comparing \texttt{col\_name} to a large integer or another column reference.

We incorporate random literal generators as callable functions within the code fragments.
Table~\ref{tab:generators} shows examples of interface functions.
They are designed to generate random literals---integers, strings, or identifiers.
\revise[R2.O4]{%
We follow the general methodology established by \ProjectName{} to bias the random generators toward fuzzing boundary values, which are more likely to trigger logic bugs.
These values include integer values near \texttt{MAX\_INT}, \texttt{MIN\_INT}, and \texttt{0}; empty, long, and Unicode strings; extreme time values.
We introduce the schema-identifier generators \texttt{RANDOM\_TABLE} and \texttt{RANDOM\_COLUMN} so that LLM-synthesized fragments can reference valid schema objects at test time.%
}
When \SystemName{}'s generator incorporates new fragments that include these keywords, it automatically invokes the matching random literal generators, dynamically generating randomized values to be included in the generated SQL statements.
By prompting LLMs with literal generators, one potential fragment could be \texttt{CHECK (col\_name > <RANDOM\_INT>)}. 
During testing, the literal generator will be invoked, and a comparison with a random integer will be generated.

\subsection{Sketch Validation}\label{sec:sketch-validation}
In step \textcircled{3}, \SystemName{} validates the synthesized code fragments using execution feedback from the target systems.
Due to LLM hallucinations, unsupported SQL features may be returned. To tackle this, we propose a validation mechanism.
Otherwise, even a small number of invalid features would result in many invalid test cases, rendering the testing process inefficient.
\revise[R1.O1 R3.Q2]{Sketch validation determines only whether a filled sketch is accepted by the target DBMS; semantic equivalence between the query pairs that the test oracle compares is enforced by the test oracle itself (\emph{e.g.}, TLP~\cite{rigger2020finding}).}

\SystemName{} validates the SQL fragments by executing the filled statements on DBMSs.
\SystemName{} substitutes the hole in the sketch with the returned fragment and executes the filled statements to check whether the features learned are supported (see step~\textcircled{3}).
After filling the holes, the statements are expected to be a self-contained SQL test case, that is, given an initial clean database state, we can execute this test case to create a table, insert data, and query it without error.
If execution of the statements is successful, we can infer that the fragments filled in have a high probability of being supported by the target system. 
Otherwise, the fragments filled in are marked as unsupported.
In the following, we refer to a fragment as \emph{valid} if it executes successfully when substituted into the sketch, and as \emph{invalid} otherwise.
For example, the unequal operator \texttt{<>} is valid, and the logical \texttt{AND} operator is invalid as shown in step~\textcircled{3}.
Note that such fragments may contain features such as functions for random data generation that pass the test in rare cases, while actually not being supported. We address this challenge during test execution (see Section~\ref{sec:fotest}).
By integrating the fragments that contain no unsupported features, \SystemName{} ensures a high validity rate for test cases.

\begin{algorithm}[tb]
    \caption{Feature-oriented testing in \SystemName{}.}
\label{alg:fuzzing}
\SetKwFunction{initializeTypes}{initializeTypes}
\SetKwFunction{generateDatabase}{generateDatabase}
\SetKwFunction{generateQueries}{generateQueries}
\SetKwFunction{oracleCheck}{oracleCheck}
\SetKwFunction{nextType}{nextType}
\SetKwFunction{focusedTesting}{focusedTesting}
\SetKwFunction{synthesize}{synthesize}
\SetKwFunction{rag}{rag}
\SetKwFunction{generateSearchQuery}{generateSearchQuery}
\SetKwFunction{searchFetch}{searchAndFetchTop}
\SetKwFunction{summarize}{summarize}
\SetKwProg{Fn}{Function}{:}{}

\SetCommentSty{mycommfont}

\Repeat{timeout}{
$\mathcal{P}$ $\leftarrow$ \initializeTypes{$D$}\;\label{algoline:initialize}
$\mathcal{L} \leftarrow \emptyset$\;
\Repeat{allTypesLearned}{
     \tcp{Pick next type $\tau\in\mathcal{P}\setminus\mathcal{L}$}
     $\tau \leftarrow$ \nextType{$\mathcal{P}, \mathcal{L}$}\;\label{algoline:nexttype}
     \ifshowrevisions{\color{blue}\tcp{Documentation summary via RAG (Section~\ref{sec:fragmentsyn})}}\else\tcp{Documentation summary via RAG (Section~\ref{sec:fragmentsyn})}\fi
     \ifshowrevisions{\color{blue}$\mathcal{R} \leftarrow$ \rag{$D, \tau, \texttt{datatype}$}}\else$\mathcal{R} \leftarrow$ \rag{$D, \tau, \texttt{datatype}$}\fi\;\label{algoline:rag}
     \tcp{Synthesize fragments for $\tau$\ifshowrevisions{\color{blue}\ using $\mathcal{R}$}\fi}
     $\mathcal{N} \leftarrow$ \synthesize{$\tau$\ifshowrevisions{\color{blue}$, \mathcal{R}$}\fi}\;\label{algoline:synthesize}
     \While{curQueries $<$ maxQueries}{\label{algoline:teststart}
        \tcp{Biased generation toward $\mathcal{N}$}
        $\mathcal{D} \leftarrow$ \generateDatabase{$\mathcal{N}$}\;
        $\mathcal{Q} \leftarrow$ \generateQueries{$\mathcal{D}, \mathcal{N}$}\;
        \oracleCheck{$\mathcal{D}, \mathcal{Q}$}\;\label{algoline:testend}
     }
     $\mathcal{L} \leftarrow \mathcal{L}\cup\{\tau\}$\;
}
}
\ifshowrevisions{\color{blue}%
\BlankLine
\Fn{\rag{$D, f, \ell$}}{\label{algoline:ragstart}
   $q \leftarrow$ \generateSearchQuery{$D, f, \ell$}\;
   $\mathcal{P}_{\text{doc}} \leftarrow$ \searchFetch{$q$}\;
   $\mathcal{R}: \langle \textit{name}, \textit{desc}, \textit{example} \rangle \leftarrow$ \summarize{$\mathcal{P}_{\text{doc}}$}\;
   \Return $\mathcal{R}$\;\label{algoline:ragend}
}}\else
\BlankLine
\Fn{\rag{$D, f, \ell$}}{\label{algoline:ragstart}
   $q \leftarrow$ \generateSearchQuery{$D, f, \ell$}\;
   $\mathcal{P}_{\text{doc}} \leftarrow$ \searchFetch{$q$}\;
   $\mathcal{R}: \langle \textit{name}, \textit{desc}, \textit{example} \rangle \leftarrow$ \summarize{$\mathcal{P}_{\text{doc}}$}\;
   \Return $\mathcal{R}$\;\label{algoline:ragend}
}\fi

\end{algorithm}

\subsection{Feature Scheduling}
\label{sec:feature-scheduling}
\SystemName{} uses a feature scheduler to increase the likelihood of using newly learned fragments in the generated test cases.
This addresses multiple issues.
First, as the fragment set grows, it is increasingly unlikely that newly learned fragments will be sampled and incorporated into generated statements, which delays bug discovery in those new features.
Besides, SQL features exhibit cross-level dependencies (Section~\ref{sec:sqlancer++}): expression-level features (\emph{e.g.}, functions and operators) require prerequisite data types to be available in the generator. For example, a function like \texttt{ARRAY\_CONCAT()} can only be meaningfully tested once the \texttt{ARRAY} data type has been learned.

Algorithm~\ref{alg:fuzzing} shows the feature scheduler.
It maintains a pool $\mathcal{P}$ of candidate dialect-specific data types and a set $\mathcal{L}$ containing all learned data types, where $\mathcal{P}$ is initialized by prompting an LLM to synthesize the DBMS's supported data types through data type sketches.
At each iteration (see line~\ref{algoline:nexttype} in Algorithm~\ref{alg:fuzzing}), \ApproachAcronym{} selects the next unseen type $\tau\in\mathcal{P}\setminus\mathcal{L}$ and triggers a dedicated learning phase that synthesizes type-level fragments, including constants of $\tau$, and related expression-level fragments that consume or produce $\tau$ (see line~\ref{algoline:synthesize} in Algorithm~\ref{alg:fuzzing}).
After integrating the validated fragments, \ApproachAcronym{} enters a type-focused testing phase of $maxQueries$ queries (see lines~\ref{algoline:teststart}--\ref{algoline:testend} in Algorithm~\ref{alg:fuzzing}), during which it temporarily biases generation toward using the newly integrated fragments, for example by creating tables and views with columns of type $\tau$ and generating expressions over $\tau$.
Once this focused phase completes, \ApproachAcronym{} marks $\tau$ as learned by adding it to $\mathcal{L}$ and repeats the process by selecting the next unseen type; when $\mathcal{P}\setminus\mathcal{L}$ is empty, \ApproachAcronym{} re-initializes $\mathcal{P}$ and removes the already-learned types.

Suppose $\mathcal{P}=\{\texttt{JSON},\texttt{ARRAY},\texttt{DECIMAL}\}$ and $\mathcal{L}=\emptyset$. 
In one iteration, \ApproachAcronym{} selects $\tau=\texttt{ARRAY}$ and runs the dedicated learning phase to synthesize type-level fragments (\emph{e.g.}, \texttt{ARRAY} constants) as well as expression-level fragments that consume/produce \texttt{ARRAY} (\emph{e.g.}, functions such as \texttt{ARRAY\_UNIQUE()} and \texttt{ARRAY\_SUM()}). 
It then enters the type-focused testing phase and biases generation toward using the new fragments, for example, by creating tables or views with \texttt{ARRAY} columns and inserting learned constant values into those columns.

\subsection{Testing}
\label{sec:fotest}
In step~\textcircled{5}, \ApproachAcronym{} generates test cases and applies the test oracles. 
Initially, it does so by using only manually implemented features. 
It subsequently expands the set of features (see step~\textcircled{4}) by executing the learning phase.
The learning incurs negligible overhead since the learning is decoupled from the test case generation as a separate thread, leaving the overall testing throughput essentially unchanged.
During testing, \SystemName{} further validates the fragments at run time.

\paragraph{Run-time validation}
\SystemName{} further validates learned fragments through a feedback mechanism during testing.
This is because, during the self-validation in step~\textcircled{3}, invalid fragments that fail non-deterministically can falsely pass the sketch validation.
For example, the \texttt{<RANDOM\_VARCHAR>} generator can generate a string \texttt{`0'}, which can be implicitly converted to an integer, meaning that we could incorrectly infer that, for example, an operator expecting only integer operands could process characters.
To address this, we reuse the same Bayesian model in \emph{SQLancer++}~\cite{zhong2025scaling} to infer whether a feature is supported at run time.
First, after each feature has been integrated, \SystemName{} generates test cases containing these features.
Second, \SystemName{} calculates the estimated probability of each feature being executed successfully based on a statistical model~\cite{gelman1995bayesian}.
Third, \SystemName{} ranks the features by their estimated probability and omits the features with low probability.
\revise[R1.O1 R3.Q2]{This Bayesian feedback estimates only the probability of feature support---that is, whether executing a feature is accepted by the DBMS---and does not validate semantic equivalence between queries.
The correctness of the generated test cases is still enforced by the test oracle.}%

\section{Implementation}~\label{sec:implementation}
We introduce the implementation details of how \SystemName{} generates SQL sketches and integrates fragments into the generator.

\paragraph{Generator instrumentation}
We modified the manually written generator to generate SQL sketches for features at four levels, following the sketch rules in Table~\ref{table:sketches}.
Specifically, we override the string construction logic of the generator and treat selected positions as \emph{holes} that can be instantiated differently between learning and testing.
In learning, the hole generates placeholders at these sites to form a sketch instance; in testing, it substitutes each placeholder with a candidate sampled from the corresponding validated fragment set.
For example, in Table~\ref{table:sketches} (row~2), we insert a hole after \texttt{CREATE TABLE}: during learning, the hole becomes a placeholder, while during testing, it can be filled with \texttt{IF NOT EXISTS}.
\revise[R2.O2]{%
Each hole is introduced by overloading the generator's \texttt{append} method: a base call \texttt{append(default)} is replaced with \texttt{append(default, holeId)}, where \texttt{default} is the original fragment and \texttt{holeId} uniquely identifies the call site and serves as the key to bind the hole to its corresponding fragment pool.
A global flag \texttt{isLearn} switches the method between learning and testing modes, which implements the two-phase process of Figure~\ref{fig:mta}: during learning, it generates a numbered placeholder \texttt{\{0\}}, \texttt{\{1\}},~\ldots{}; during testing, it samples a validated fragment from the pool keyed by \texttt{holeId}.
Listing~\ref{listing:instrumentation} shows an excerpt at a function hole site, with the original line marked as removed and the instrumented call as added.
}
\revise{\begin{figure}[tb]
\captionof{lstlisting}{Example excerpt of the generator instrumentation.}
\label{listing:instrumentation}
\centering
\begin{minipage}{.52\columnwidth}
\lstset{language=Java, xleftmargin=1em, numbersep=4pt, lineskip=0pt, aboveskip=0pt, belowskip=0pt, basicstyle=\ttfamily\footnotesize, numbers=left, numberstyle=\footnotesize, commentstyle=\color{javagreen}, keywordstyle=\color{black}\bfseries, breaklines=true, breakatwhitespace=false, columns=fullflexible}
\begin{lstlisting}[firstline=1, lastline=1, backgroundcolor=\color{red!30}]
- sb.append("SIN");
\end{lstlisting}
\begin{lstlisting}[firstline=2, lastline=2, firstnumber=2, backgroundcolor=\color{green!30}]
+ sb.append("SIN", Hole.FUNC);
\end{lstlisting}
\begin{lstlisting}[firstline=3, lastline=7, firstnumber=3]

void append(String def, Hole holeId) {  // overloaded
  if (isLearn) sb.append("{" + nextHoleId() + "}");
  else sb.append(fragmentPool.get(holeId).sample());
}
\end{lstlisting}
\end{minipage}
\end{figure}
}

\paragraph{Fragment materialization}
Fragments contain \emph{dynamic components} that depend on the current database state, including schema object references (\emph{e.g.}, tables/columns) and random literal generators.
To keep learned fragments reusable across schemas and executions, we represent such components using \emph{unique identifiers} with a fixed name and description exposed to the LLM during learning (\emph{e.g.}, \texttt{RANDOM\_INT}: generate a random integer value).
During testing, the generator materializes these identifiers into concrete values: it rebinds schema identifiers to valid objects in the current schema, and it evaluates random-generator identifiers into constants using a seeded generator.

\section{Evaluation}
We evaluated \SystemName{} from multiple perspectives.
First, we evaluated the effectiveness of \SystemName{} in finding bugs over a long-term testing campaign, and analyzed the bugs we found from different levels of features (see Section~\ref{sec:bugs}).
Second, we compared \SystemName{}---under different configurations---with state-of-the-art logic bug detection tools (see Section~\ref{sec:coverage}).
Third, we measured the contribution of each component in \SystemName{} to bug finding effectiveness (see Section~\ref{sec:ablation}).

\begin{table*}[t]
    \centering
\small
\caption{\SystemName{} allowed us to find and report \OverallBugs{} bugs in \DBMSUnderTestNum{} systems, of which \FixedBugs{} were fixed by the developers, and \LogicBugs{} were logic bugs. In terms of bug types, 8 were attributed to clause features, 26 to type features, and 21 to expression features. Notably, none of these bugs could be detected by SQLancer++~\cite{zhong2025scaling}.}\label{table:DBMSs}
\resizebox{\columnwidth}{!}{%
\begin{tabular}{ lrrrrrrrrrrrlr }
\toprule
DBMS & \multirow{2}{*}{All} & \multicolumn{2}{c}{Bug Status} & \multicolumn{2}{c}{Test Oracle} & \multicolumn{3}{c}{SQL Feature} & \multirow{2}{*}{Released} & \multirow{2}{*}{LOC} & \multirow{2}{*}{\shortstack{GitHub\\Stars}} & \multirow{2}{*}{\shortstack{Source\\Language}} & \multirow{2}{*}{\shortstack{Prior\\Tools}} \\
Names & & Fixed & Conf. & Logic & Other & Clause & Type & Expr. & & & & & \\
\midrule
CockroachDB &  2 &  1 & 1 &  1 &  1 &  0 &  1 &  1 & 2014 & 2522k & 31.7k & Go & \cite{zhong2025scaling,rigger2020detecting,ba2023testing,rigger2020finding,song2024detecting} \\
CrateDB     & 21 & 21 & 0 & 20 &  1 &  7 &  8 &  6 & 2017 & 563k & 4.3k & Java &  \cite{zhong2025scaling} \\
DuckDB      & 16 & 15 & 1 & 13 &  3 &  0 & 15 &  1 & 2019 & 721k & 35.2k & C++ & \cite{fu2024sedar,fu2023griffin,zhong2025scaling,rigger2020finding,zhang2025constant} \\
MonetDB     & 11 & 11 & 0 &  2 &  9 &  1 &  1 &  9 & 2004 & 339k & 0.5k & C &  \cite{fu2024sedar,zhong2025scaling} \\
TiDB        &  5 &  2 & 3 &  3 &  2 &  0 &  1 &  4 & 2016 & 1212k & 39.6k & Go & \cite{zhong2025scaling,ba2023testing,zhang2025constant,song2023testing,song2024detecting,tang2023detecting,rigger2020finding} \\
\midrule
\textbf{Total} & 55 & 50 & 5 & 39 & 16 &  8 & 26 & 21 & \multicolumn{5}{c}{ } \\
\bottomrule
\end{tabular}
}

\end{table*}
\paragraph{Baselines}
We implemented \SystemName{} based on \ProjectName{}. 
It consists of 10.5K LOC written in Java for the base generator and validator, and 200 LOC of Python for building LLM agents.
In comparison, SQLancer has 83K LOC and \emph{SQLancer++} has 8.4K LOC.
We compared \SystemName{} with \emph{SQLancer++} and SQLancer.
Our aim for \SystemName{} was to detect bugs in various DBMSs with different dialects.
In addition, we sought to make it more effective and support more features than \emph{SQLancer++}.
We did not aim to outperform SQLancer's manually written, dialect-specific generators; rather, our goal was to show the feasibility of fully automatic bug detection for dialect-specific features.

\paragraph{Setup}
We conducted the experiments using a server with a 64-core EPYC 7763 at 2.45 GHz and 512 GB of memory running on Ubuntu 22.04.
In \SystemName{}, we used the OpenAI APIs to query LLMs. Specifically, we used GPT-4o for SQL sketch synthesis and GPT-4o-mini for document summarization. 
During each testing iteration, we randomly created up to 2 tables, 1 view, and 20 inserts, and then executed 100K queries. These are the standard settings for SQLancer and SQLancer++.
We also compared \SystemName{} under different LLM settings.
``\textit{ShQveL\textsubscript{Model}}'' denotes using only the LLM's internal knowledge without DBMS documentation.

\paragraph{DBMS selection}
We evaluated the bug-finding effectiveness of our approach on \DBMSUnderTestNum{} DBMSs---CockroachDB, CrateDB, DuckDB, MonetDB, and TiDB (see Table~\ref{table:DBMSs}). 
We used the latest development versions and reported bugs only when they could be reproduced on their latest versions.
SQLancer provides support for all DBMSs except CrateDB, which is supported only by SQLancer++.
Among the systems, CockroachDB, DuckDB, MonetDB, and TiDB have previously been the focus of both logic‑bug~\cite{song2024detecting, liu2024semantic} and memory-bug~\cite{fu2024sedar} detection techniques.
We also selected these five systems because their development communities actively investigate and resolve reported issues, enabling us to validate the uniqueness of any bugs we uncover.
Although we also detected bugs in other popular DBMSs (\emph{e.g.}, MySQL and MariaDB), many previously reported bugs remain unfixed, making it challenging to determine whether any bug-inducing test cases trigger known bugs~\cite{song2024detecting, hao2023pinolo, rigger2020testing}.
To avoid overloading the developers, we refrained from reporting any additional bugs for these systems.
In our experiments on code coverage evaluation (see Section~\ref{sec:coverage}), we compared \SystemName{} with the baseline tools---\SQLancer{}, \ProjectName{}, and SQLRight---on DuckDB, SQLite, and PostgreSQL, which differs from the DBMSs used for evaluating bug-finding effectiveness, since AFL-based tools (\emph{e.g.}, SQLRight~\cite{liang2022detecting}) cannot be applied to non-C/C++ systems---CockroachDB, CrateDB, and TiDB---and we can still trigger crashes on the latest version of MonetDB, which would cause any experiments to prematurely terminate.

\subsection{Bug Finding Effectiveness}\label{sec:bugs}
We continuously tested the five DBMSs during a six-month fuzzing campaign, followed by several months of intermittent testing.
This methodology was also used to evaluate various other automated testing approaches, such as SQLancer++~\cite{zhong2025scaling} and other testing approaches for DBMSs~\cite{jiang2023detecting, rigger2020testing, ba2023testing}.
Because existing approaches have exhaustively tested the target systems~\cite{fu2024sedar, ba2023testing}, any new bugs found by \SystemName{} indicate cases that existing tools missed.
We ran \SystemName{} for several minutes to one day until it generated a number of bug reports. 
We employed the bug reduction and prioritization mechanisms of SQLancer++, and then we further processed the reduced and prioritized bug reports.
After the bugs were fixed, we updated the DBMS to the latest version and started another testing run on it.

\paragraph{Bug statistics}
Table~\ref{table:DBMSs} demonstrates the statistics of the bugs we reported. The full bug list is included in the artifact.
In total, we reported \OverallBugs{} bugs, of which \FixedBugs{} bugs have been fixed, and the rest have been confirmed.
\LogicBugs{} out of the \OverallBugs{} reported bugs were logic bugs, including \LogicBugCockroachDB{} bug in CockroachDB, \LogicBugCrateDB{} bugs in CrateDB, \LogicBugDuckDB{} bugs in DuckDB, \LogicBugMonetDB{} bugs in MonetDB and \LogicBugTiDB{} bugs in TiDB. 
\SystemName{} also found \OtherBugs{} other bugs causing hangs, internal errors, or crashes in the DBMS.
This is expected, since automatically incorporating new features could help \SystemName{} detect issues in these previously untested or not well-tested features.
\SystemName{} detected the logic bugs by the TLP~\cite{rigger2020finding} oracle, and detected other bugs (\emph{e.g.}, crash and hang) by monitoring the DBMS process status and response latency.
These results are encouraging and demonstrate that \SystemName{} can find bugs that the DBMS developers are willing to fix.

\paragraph{Feature analysis}
~\label{sec:bug-studies}
Table~\ref{table:DBMSs} presents bugs found by \SystemName{} augmented with SQL features at different levels.
Most of the bugs found on DuckDB are caused by data type features, including \texttt{BIT}, \texttt{TIMEZONE}, and \texttt{INET}. 
These data types are not supported by SQLancer, which is why \SystemName{} could successfully find them. 
Most bugs in TiDB and MonetDB stem from expression-level features---functions and operators.
Extending existing generators with new functions or operators might require less effort than adding clause- or statement-level features; however, manually covering hundreds of built-in or newly released features is impractical.
\SystemName{} can automatically incorporate these new features.

Subsequently, we present the bugs we found at different feature levels.
Our goal is to show the breadth of distinct bugs identified by each feature we learned. 
We present only reduced test cases that still contain the feature and highlight the core issue, rather than the original and semantically equivalent queries used to uncover these bugs.

\begin{lstlisting}[language=SQL,float=tb,  caption={A bug in TiDB when using a JSON function.}, label={listing:function-bug}, escapechar=@, 
showspaces=false,
numbers=left,
xleftmargin=1em,
xrightmargin=-0.5cm,
numberstyle=\scriptsize,
commentstyle=\color{gray},
keywordstyle=\color{black}\bfseries,
breaklines=true,
basicstyle=\ttfamily\footnotesize,
breakatwhitespace]
CREATE TABLE t0(c0 VARCHAR(500), c1 INT);
CREATE VIEW v0(c0) AS SELECT 'a' FROM t0;
INSERT INTO t0(c0) VALUES ('b');

SELECT * FROM t0 NATURAL RIGHT JOIN v0 
    WHERE (JSON_VALID(t0.c1)=0);
-- {} @\faBug@ -- {'a' NULL} @\faCheck@
\end{lstlisting}

\paragraph{Incorrect function result}
Listing~\ref{listing:function-bug} demonstrates a bug in TiDB that was related to function-level features.
The TiDB developers explained it was an inconsistent behavior for the \texttt{JSON\_VALID} function when its argument is a column reference or a literal \texttt{NULL}. 
Surprisingly, the developers discovered that MySQL was also affected by this bug, and thus, they submitted an issue report to the MySQL developers.
This bug was found by the TLP oracle; however, \SQLancer{} failed to find it, since it lacked support for this function.

\begin{lstlisting}[language=SQL,float=tb,  caption={A large integer argument led to a CrateDB crash.}, label={listing:clause-crash}, escapechar=@,
showspaces=false,
numbers=left,
xleftmargin=1em,
xrightmargin=-0.5cm,
numberstyle=\scriptsize,
commentstyle=\color{gray},
keywordstyle=\color{black}\bfseries,
breaklines=true,
basicstyle=\ttfamily\footnotesize,
breakatwhitespace]
CREATE TABLE t0(c0 INT) WITH (number_of_replicas = 1000);
-- raise a SQLParseException
CREATE TABLE t0(c0 INT) WITH (number_of_replicas = 1925152226);
-- hang and crash
\end{lstlisting}

\paragraph{Table property clause crash}
We found a bug in CrateDB (see Listing~\ref{listing:clause-crash}) in which a \texttt{CREATE TABLE} statement that set a large integer value as a table constraint led to a hang and crash of the system.
The system crashed due to an integer overflow when validating the shards limit; however, CrateDB failed to catch the exception.
The \texttt{WITH} clause is not defined in standard SQL and the attribute \texttt{number\_of\_replicas} is CrateDB-specific.
\ApproachAcronym{} learned this statement-level fragment and used it while creating tables, which led to the crash.
Although \SystemName{} was not explicitly designed for detecting crash bugs, our experiments demonstrate its capability to uncover several such issues.
This bug also shows the benefit of the built-in random literal generator, since the LLM trained on the real-world dataset is unlikely to directly generate the code fragments with an overly large integer value.

\begin{lstlisting}[language=SQL,float=tb,  caption={Column constraints cause CrateDB to incorrectly select a hash join plan.}, label={listing:clause-logic}, escapechar=@, 
showspaces=false,
numbers=left,
xleftmargin=1em,
xrightmargin=-0.5cm,
numberstyle=\scriptsize,
commentstyle=\color{gray},
keywordstyle=\color{black}\bfseries,
breaklines=true,
basicstyle=\ttfamily\footnotesize,
breakatwhitespace]
CREATE TABLE t0 (c1 INT NOT NULL DEFAULT 1);
CREATE TABLE t1 (c1 INT);
INSERT INTO t0(c1) VALUES (1);
INSERT INTO t1(c1) VALUES (2);

SELECT * FROM t0, t1 WHERE
  (t1.c1 >= 1) =((t1.c1 = t1.c1)AND(t0.c1 <= t0.c1));
-- {} @\faBug@ -- {1 2} @\faCheck@
\end{lstlisting}

\paragraph{False optimization by default clause}
Listing~\ref{listing:clause-logic} shows a bug caused by incorrectly selecting a hash join plan.
As explained by the developers, the issue arises when joining tables whose filter involves multiple nested equality comparisons, and manifests only when the column constraints \texttt{NOT NULL DEFAULT 1} are enabled.
\ApproachAcronym{} learns the clause-level fragments and can generate a database schema with DBMS-specific features, which may trigger potential optimizations.

\begin{lstlisting}[language=SQL,float=tb,  caption={DuckDB incorrectly evaluated a cast of a time value.}, label={listing:datatype-bug}, escapechar=@,
showspaces=false,
numbers=left,
xleftmargin=1em,
xrightmargin=-0.5cm,
numberstyle=\scriptsize,
commentstyle=\color{gray},
keywordstyle=\color{black}\bfseries,
breaklines=true,
basicstyle=\ttfamily\footnotesize,
breakatwhitespace]
CREATE TABLE t1(c0 TIME WITH TIME ZONE);
INSERT INTO t1(c0) VALUES ('12:34:56');

SELECT (CAST(t1.c0 AS TIME) IN ('12:34:56')) FROM t1;
-- false @\faBug@ -- true @\faCheck@
\end{lstlisting}

\paragraph{Data type}
We found a bug in DuckDB (see Listing~\ref{listing:datatype-bug}) where an \texttt{IN} expression was unexpectedly evaluated to \texttt{false}.
We found this bug by the TLP oracle for which none of the partitioning queries of this expression predicate fetched any rows.
As explained by the developers, this was due to a bug in the implementation of cast for \texttt{TIME} data types. They also mentioned that the support for this data type is not well-tested.\footnote{https://github.com/duckdb/duckdb/issues/13813}
Manually implementing these features into existing generators is labor-intensive.
By automatically learning from an LLM, \ApproachAcronym{} can easily uncover untested features and trigger potential bugs.

\begin{table}[tb]
    \centering
\small
\setlength{\tabcolsep}{4pt}
\caption{Line and branch coverage on SQLite, PostgreSQL, and DuckDB after 24~hours, comparing \SystemName{} (with and without documentation summarization) against \ProjectName{}, \emph{SQLRight}, \SQLancer{}, and Argus.}
\label{tab:coverage}
\begin{tabular}{lrrrrrr}
\toprule
\multirow{2}{*}{Approach}       & \multicolumn{2}{c}{SQLite}&  \multicolumn{2}{c}{PostgreSQL}&  \multicolumn{2}{c}{DuckDB}  \\
                                & Line  & Branch & Line  & Branch   & Line  & Branch    \\
\midrule
\textit{ShQveL}  &  41.3\%  & 33.2\% &  31.6\% &  22.9\%  &  44.6\% & 27.2\% \\
\textit{ShQveL\textsubscript{Model}}    & 38.2\%  &  30.6\% & 30.4\% &  21.9\% & 41.7\% &  24.9\% \\
\textit{SQLancer++}        &  30.3\%   &  23.0\% &  25.4\% &  17.8\%  &  31.6\% & 18.8\% \\
\textit{SQLRight}    &  47.5\%   &  37.7\% & 44.9\% &  34.2\%  &   - & - \\
\textit{SQLancer}    & 46.6\%  & 36.4\% & 32.3\% &  23.3\%  &  33.4\% & 20.9\%\\
\textit{Argus}     & --        & --     & 30.1\%     & 21.8\%       & 37.1\%     & 21.6\% \\

\bottomrule
\end{tabular}

\end{table}

\subsection{Baseline Comparison}~\label{sec:coverage}
We measured the performance using multiple configurations of \ApproachAcronym{} and baselines.
We compared the bug detection efficiency between \emph{ShQveL} and \emph{SQLancer++} on CrateDB 5.6, a historical version. Using a historical version for evaluating the efficiency is common practice~\cite{zhang2025constant, song2024detecting}, as the uniqueness of bugs can be determined by identifying which commit fixed a bug and clustering all bug-inducing test cases by the commit that fixed them.
We also evaluated the code coverage of \ApproachAcronym{} under different settings, SQLancer, SQLRight, SQLancer++, and Argus where applicable on three DBMSs written in C/C++: SQLite, PostgreSQL, and DuckDB.
Although code coverage is not a crucial metric to measure the capabilities of finding logic bugs, it can help relatively compare how many features have been covered.
We do not expect \SystemName{} to outperform SQLancer or SQLRight, since they require manually written generators or parsers, which are specific to each system, while \SystemName{} can be easily adapted to SQL dialects that are not supported by existing tools.

\subsubsection{Coverage}\label{sec:coverage-cov}

\begin{figure*}
    \centering
    \includegraphics[width=0.95\linewidth]{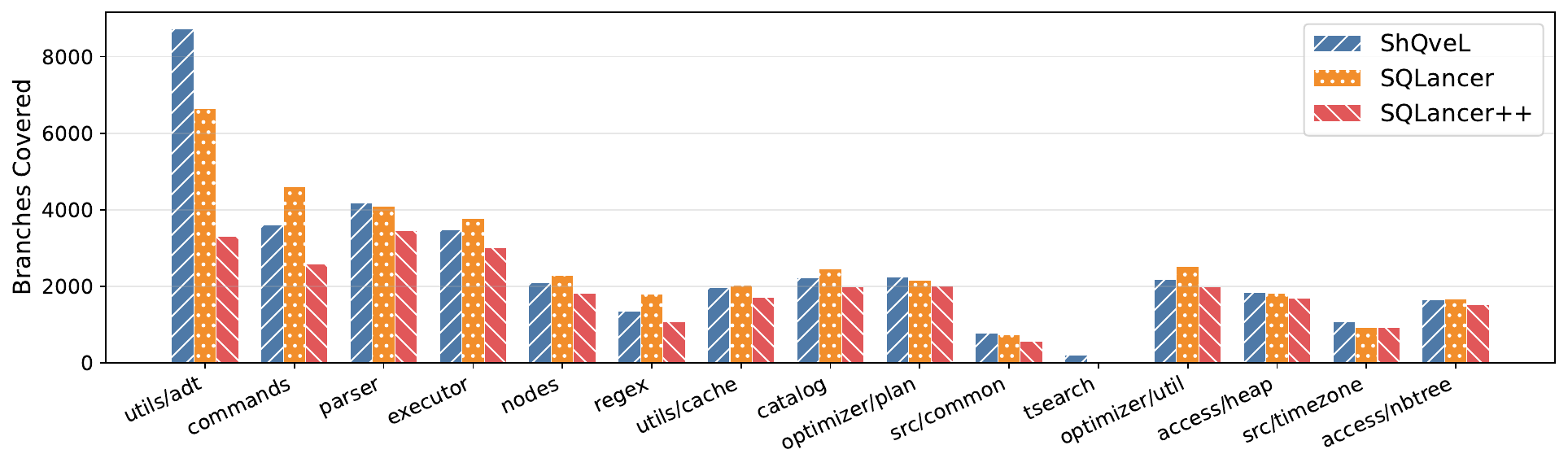}
    \caption{Branches covered in PostgreSQL by top-level directory (from the \texttt{lcov} report), comparing \SystemName{}, \SQLancer{}, and \ProjectName{}. We show the 15 directories with the largest coverage difference between \SystemName{} and \ProjectName{}.}
    \label{fig:pg-branch-coverage}
\end{figure*}

\paragraph{Code coverage}
Table~\ref{tab:coverage} demonstrates the average line and branch coverage across 10 runs in 24 hours, which adheres to best practices~\cite{klees2018evaluating}.
\emph{SQLRight}, which is based on \emph{Squirrel}, achieves the highest coverage since it was designed to maximize code coverage. Besides, the coverage of initial seed inputs from each DBMS test suite already exceeds that of \emph{SQLancer}.
\emph{SQLancer} achieves relatively higher coverage on PostgreSQL and SQLite.
For \emph{SQLancer} and \emph{SQLRight}, significant manual effort was necessary (\emph{e.g.}, 9.7K LOC for the SQLite generator in SQLancer and an average of 7.9K LOC for \emph{Squirrel}, on which SQLRight is based) to support these features, whereas \SystemName{} can achieve similar performance automatically.
Note that on DuckDB, \SystemName{} outperformed SQLancer.
One plausible explanation is that DuckDB, as an emerging and popular DBMS, evolves rapidly with new features, whereas SQLancer may not incorporate them as quickly.
By incorporating LLM-derived features, \emph{ShQveL} increases branch coverage over \emph{SQLancer++} by 44.3\% on SQLite, 28.7\% on PostgreSQL, and 44.7\% on DuckDB.
Furthermore, the documentation summarization in \emph{ShQveL} achieves 5\%-9\% more coverage than using only the LLM's internal knowledge.
\revise[R2.O1]{\SystemName{} also achieves higher branch coverage than Argus~\cite{mang2025automated}, a concurrent LLM-based DBMS-testing approach, on both PostgreSQL and DuckDB.}

\paragraph{Fine-grained coverage analysis}
Figure~\ref{fig:pg-branch-coverage} provides a finer-grained comparison of branch coverage in PostgreSQL across \SystemName{}, \SQLancer{}, and \ProjectName{} in different source code directories.
\SystemName{} significantly outperforms the baselines in the \texttt{utils/adt} directory---Abstract Data Type---and slightly outperforms the baselines in the parsing and planning pipeline (\emph{e.g.}, \texttt{parser},  \texttt{optimizer/plan}, and \texttt{optimizer/util}).
However, \SystemName{} is less competitive in components like \texttt{commands}, \texttt{regex}, and some downstream execution paths (\emph{e.g.}, \texttt{access/heap}).
One reason is that \SQLancer{}'s customized generators have a higher validity rate---they more often satisfy prerequisite schemas, modes, and cross-clause dependencies due to manually specified rules or heuristics---so more generated inputs survive past parsing/planning and reach these components.

We observed that \SQLancer{} can achieve higher coverage for regex-related functionality, because it includes a specialized regex expression generator. 
It reliably constructs valid patterns and generates operators or functions such as \texttt{\~}, \texttt{regexp\_match}, and \texttt{regexp\_replace} with well-formed regex patterns as inputs.
In contrast, \SystemName{} treats the regex pattern as an unconstrained string and may produce semantically invalid patterns that are rejected; learning such constraints might require a domain-specific grammar or specification for the generator.
Conversely, the higher coverage of \SystemName{} in \texttt{utils/adt} is consistent with our feature-level dialect comparison in Table~\ref{tab:feature-compare}: \SystemName{} learns a wider variety of built-in data types and, together with our random literal generator, produces more diverse constants that stress type-specific parsing, coercions, and operators (\emph{e.g.}, \texttt{NUMERIC/DECIMAL} literals, nested \texttt{ARRAY} literals, and structured \texttt{JSON/JSONB} values).

\begin{table}[tb]
    \centering
\small
\setlength{\tabcolsep}{4pt}
\caption{Feature-level dialect coverage comparison between \SQLancer{} and \SystemName{}.}
\label{tab:feature-compare}
\begin{tabular}{lrrrrrr}
\toprule
    & \multicolumn{3}{c}{\SQLancer{}} & \multicolumn{3}{c}{\SystemName{}} \\
\cmidrule(lr){2-4}\cmidrule(lr){5-7}
\multirow{-2}{*}{DBMS} & Stmt. & Datatype & Expr. & Stmt. & Datatype & Expr. \\
\midrule
SQLite     & 17 & 4  & 77  & 16 & 4  & 76  \\
DuckDB     & 8  & 6  & 114 & 31 & 13 & 157 \\
PostgreSQL & 22 & 10 & 151 & 29 & 16 & 144 \\
\bottomrule
\end{tabular}

\end{table}

\paragraph{Feature coverage}
We quantify dialect coverage by counting distinct, deduplicated \emph{features} at three levels: statements, data types, and expressions.
We exclude clause-level fragments because embedded constants or identifiers make deduplication unreliable.
For example, \SystemName{} may generate column-constraint clauses such as \texttt{CHECK {COL} > 1} and \texttt{CHECK {COL} < 1}.
These differ only in the instantiated constant and comparison operator, not in the underlying core feature (\emph{e.g.}, a \texttt{CHECK} constraint over a column predicate).
Counting them as separate clause features would therefore overcount what is essentially the same underlying clause feature.
\revise[R2.O7]{
Parsing each clause-fragment into an AST may be a potential improvement, and we leave it as future work.
}
For \SystemName{}, a feature is a validated, unique fragment learned from documentation and classified by
the corresponding sketch level.
For \SQLancer{}, which lacks an explicit list of supported features, we approximate feature sets by inspecting its dialect-specific generators and grouping the covered AST node kinds into the same three levels.

Table~\ref{tab:feature-compare} compares the resulting feature coverage on SQLite, DuckDB, and PostgreSQL.
The results show that the feature coverage of \SystemName{} is comparable to \SQLancer{} on SQLite and PostgreSQL, while achieving higher coverage at all three feature levels on DuckDB.
On DuckDB, the additional features include more statement forms (\emph{e.g.}, bulk-load/export statements like \texttt{COPY}), built-in data types (\emph{e.g.}, \texttt{JSON/JSONB}, \texttt{DECIMAL/NUMERIC}, and nested types such as \texttt{ARRAY}), and expression nodes (\emph{e.g.}, JSON extraction, array/list constructors, and time-related functions).
On PostgreSQL and SQLite, \SQLancer{} achieves higher code coverage than \SystemName{}, because \SQLancer{}'s handwritten generators are optimized to produce test cases that are less likely to be rejected by the DBMS, as described in the previous paragraph.

\subsubsection{Bug Finding Efficiency}\label{sec:coverage-bugs}

\begin{figure}
    \centering
    \includegraphics[width=0.65\linewidth]{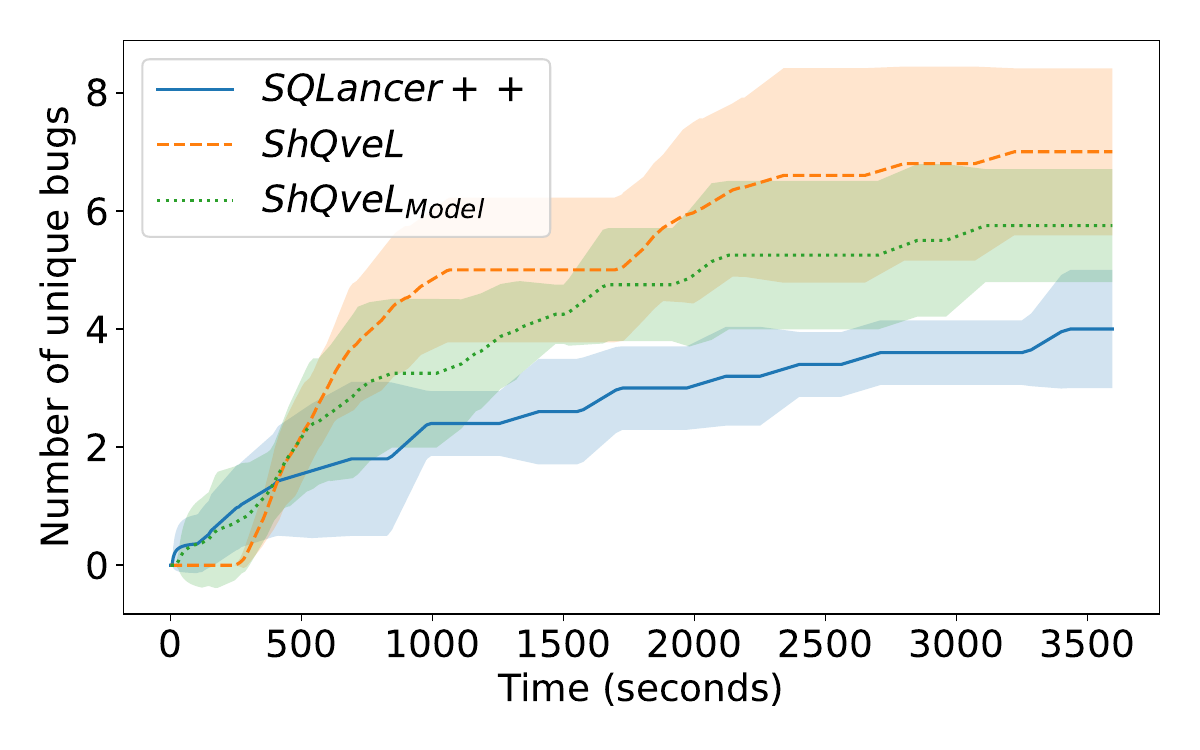}
    \caption{\revise[R2.O6]{Unique bugs found on CrateDB within one hour, averaged across 5 independent runs. The shaded area shows the standard deviation.}}
    \label{fig:crate-bug}
\end{figure}

Figure~\ref{fig:crate-bug} shows the bug detection efficiency of \emph{ShQveL}, \emph{ShQveL} without reference summarization (\textit{ShQveL\textsubscript{Model}}), and SQLancer++ on CrateDB.
In one hour, \SystemName{} detected more bugs than \textit{ShQveL\textsubscript{Model}}, as it generates features described in the documentation that would otherwise be missed.
Listing~\ref{listing:example-crate} illustrates a class of bugs missed by \textit{ShQveL\textsubscript{Model}}; we found that LLMs often fail to provide dialect-specific SQL features for DBMSs that might be underrepresented in their training data.
Both \SystemName{} and \SystemName{} without summarization can outperform SQLancer++, since \SystemName{} can discover new bugs by LLM-derived features.
\SystemName{} found its first bugs slightly later than SQLancer++ because it started learning from scratch, and summarizing the documentation takes time.

\begin{table}[tb]
\centering
\small
\caption{Branch coverage achieved by \SystemName{} on SQLite, PostgreSQL, and DuckDB after 24 hours when learning exclusively from statement, clause, datatype, and expression features.}
\label{tab:coverage-incremental}
\begin{tabular}{lrrrrr}
\toprule
DBMS       & Base &  Statement &  Clause & Datatype & Expression \\
\midrule
SQLite          & 23.0\%  & 28.6\% &   24.9\% &  23.6\%  & 24.6\% \\
PostgreSQL            & 17.8\% & 20.0\%  &  18.0\% &  19.8\%  & 19.8\% \\
DuckDB     & 18.8\%  &  22.9\% &  19.2\% &  22.4\%  & 20.7\% \\

\bottomrule
\end{tabular}

\end{table}

\begin{figure}
    \centering
    \includegraphics[width=0.98\linewidth]{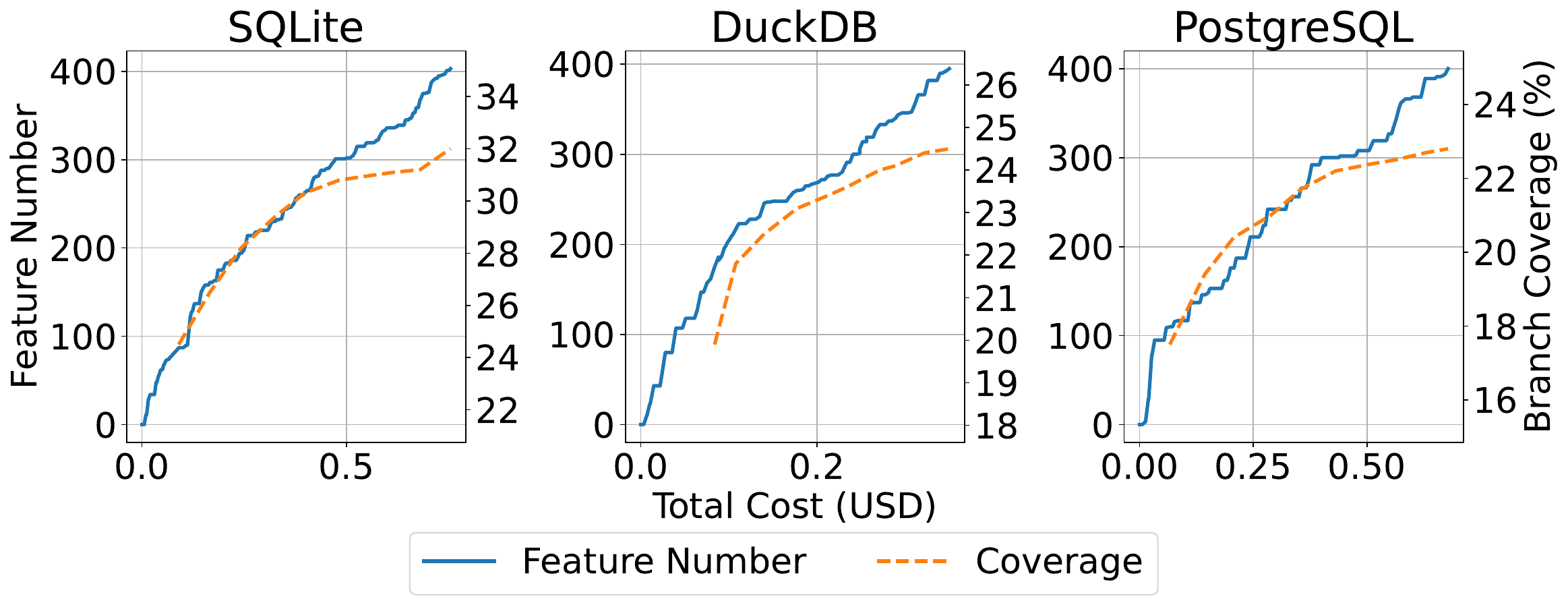}
    \caption{The cost of LLM APIs to learn features on SQLite, DuckDB, and PostgreSQL over six hours.}
    \label{fig:feature-cost}
\end{figure}

\subsection{Ablation Study}~\label{sec:ablation}
We further measured the contributions and costs of individual components of \SystemName{}.

\paragraph{Feature importance}
We measured the effectiveness of learning features at each level.
We evaluated \SystemName{} on SQLite, PostgreSQL, and DuckDB by enabling learning only one type of feature during each execution.
Table~\ref{tab:coverage-incremental} shows the incremental learning effect in terms of branch coverage of each individual level of feature. The base column represents using the base generator.
Statement-level features lead to the greatest coverage increase across all three DBMSs, since the base generator supports only common SQL statements.
\SystemName{} found no new logic bugs by learning new statement-level features, since most logic errors stem from issues in the query processor rather than the statement executor.
Expression‑level features are dependent on data types (see Section~\ref{sec:sketch-design}).
However, in our experimental setup, the functions and operators were learned without their associated data types, causing most of them to fail.

\paragraph{Feature learning costs}
We measured the cost required by \SystemName{} to discover new features and improve branch coverage over a six-hour run on SQLite, DuckDB, and PostgreSQL.
Each learning phase was triggered only after executing a sufficient number of test cases (\emph{e.g.}, 200K SQL statements).
We incremented the feature count when \SystemName{} generated a valid, previously unseen code fragment for a given level of SQL sketch, filtering out duplicates and invalid fragments.
We also recorded the branch coverage of the DBMSs during execution.
Figure~\ref{fig:feature-cost} demonstrates the cumulative number of learned features (solid line) and branch coverage (dashed line) against total API cost in USD.
\SystemName{} learns over 400 features on all three DBMSs for under 1 US dollar, achieving branch coverage close to that of the original SQLancer.
On DuckDB, it learns around 400 features for less than \$0.4 and outperforms SQLancer.
Note that SQLite’s higher throughput triggers more frequent learning phases, and thus has a higher cost.

\begin{figure}
    \centering
    \includegraphics[width=0.98\linewidth]{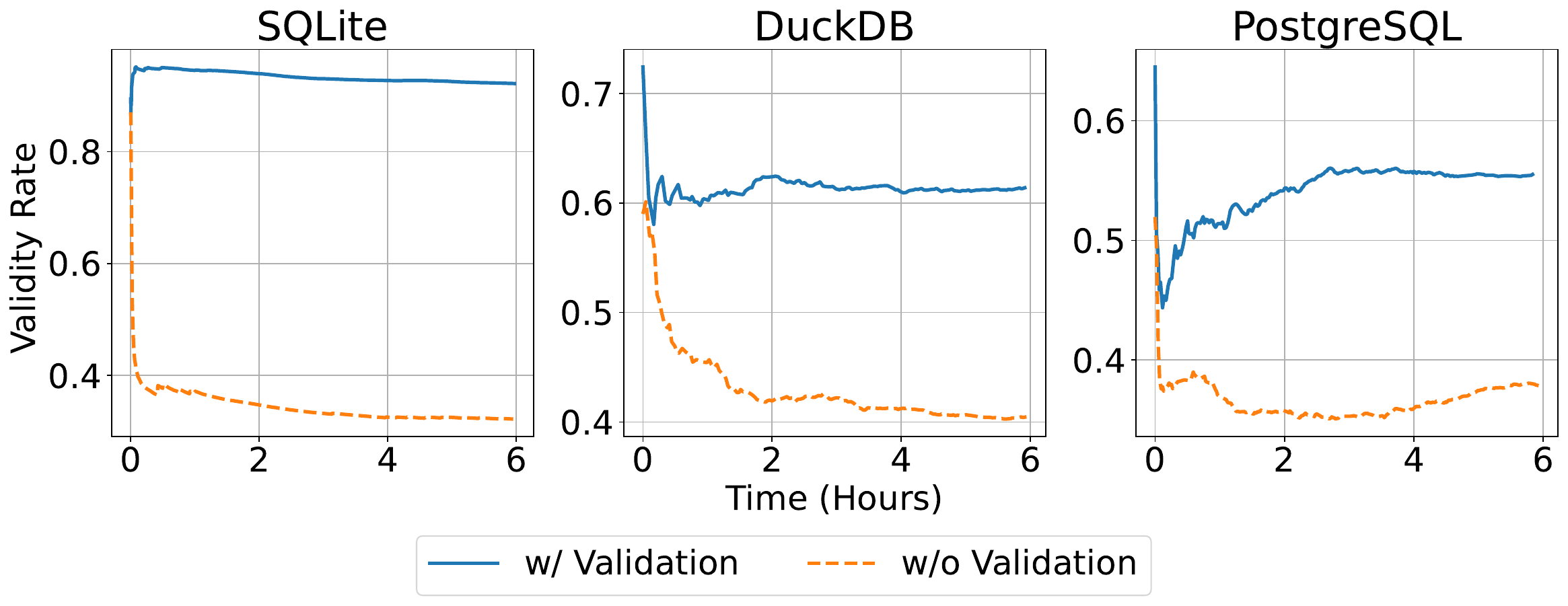}
    \caption{The cumulative validity rate of test cases generated by \SystemName{} on SQLite, DuckDB, and PostgreSQL over six hours.}
    \label{fig:validity}
\end{figure}

\paragraph{Validity rate}
We measured the query and statement validity rates of \SystemName{} when learning new features over a six-hour run on SQLite, DuckDB, and PostgreSQL with and without the fragment validation mechanisms.
Figure~\ref{fig:validity} shows the cumulative success rate of the generated SQL statements, including both DDL and DML statements, and queries over time.
When fragment validation is disabled, the validity rate declines by \SQLiteValidityDrop{}, \DuckDBValidityDrop{}, and \PostgreSQLValidityDrop{} on SQLite, DuckDB, and PostgreSQL, respectively.
The validity rate remains above zero, since \SystemName{} generates common features besides the newly learned features.
Among the three systems, SQLite achieves the highest overall validity, due to its dynamic typing and ability to coerce most values into the required types.

\revise[R1.O4]{%
\result{LLM usage is incurred once per DBMS during the initial learning phase: learning around 400 fragments costs at most \$1 per DBMS and completes in under six hours.
After that, \SystemName{} invokes no LLM during test-case generation and its throughput matches \ProjectName{}.
Learned fragments persist across runs and across DBMS versions, so the one-time cost is amortized over every subsequent CI run.}
}

\begin{table}
    \centering
    \caption{False alarms observed at each step of the incremental filter construction on PostgreSQL~17.0.}
    \label{tab:fa_curve}
    \begin{tabular}{rrrr}
        \toprule
        Step & Filter size & Run time & False alarms\\
        \midrule
        0 & 0 & 1\,h  & 17.0 \\
        1 & 10 & 1\,h  & 2.4 \\
        2 & 15 & 1\,h  & 1.4  \\
        3 & 19 & 1\,h  & 0.4  \\
        4 & 20 & 1\,h  & 0  \\
        5 & 20 & 24\,h & 0  \\
        \bottomrule
    \end{tabular}
\end{table}
\subsection{False Alarm Analysis}\label{sec:fp}
\revise[R1.O2 R2.O5]{
False alarms can arise when \SystemName{} learns fragments that violate the constraints of the test oracle (\emph{e.g.}, non-deterministic functions), leading to discrepancies in query results that may be misinterpreted as bugs.
To reduce such alarms, \SystemName{} maintains a keyword filter that excludes any learned fragment containing such features.
We sought to measure how often false alarms would arise without this filter, and the manual effort required to build it.

\paragraph{Results}
We conducted the analysis on PostgreSQL~17.0, constructing the filter incrementally to observe how the number of false alarms decreased after each addition.
At each step, we ran five \SystemName{} instances in parallel on PostgreSQL for one hour, collected all bug-inducing test cases, and manually classified them as true bugs or false alarms.
For each false alarm, we added the responsible identifier to the filter.
For the final step, in which no new entry was added, we extended the run to 24 hours to increase the chance of triggering any remaining false alarms.
Table~\ref{tab:fa_curve} reports the filter size and the average number of false alarms observed at each step.
The resulting filter covers three types of fragments: non-deterministic functions, set-returning functions, and transaction-control statements.
}

\section{Discussion}\label{sec:discussion}

\paragraph{Design rationale}
\revise[R3.O4]{%
Our evaluation demonstrates that each component of ShQveL contributes to the approach's overall effectiveness, thus justifying \SystemName{}'s design.
First, to systematically cover the SQL features, we define four hole types (statement, clause, expression, data type) corresponding to the taxonomy of prior work~\cite{zhong2025scaling}; the coverage study (see Table~\ref{tab:coverage-incremental}) shows that each level contributes incremental code coverage.
Second, to address the LLM's knowledge gap on emerging or less common DBMSs, we use a RAG-based approach to incorporate the information from the documentation of the target DBMS, which yields 5--9\% additional branch coverage (see Table~\ref{tab:coverage}).
Third, to prevent hallucinated fragments from reducing testing throughput, we filter fragments through an execution-based feedback mechanism; without it, the statement validity rate drops (see Figure~\ref{fig:validity}), making testing inefficient.
While each component adds implementation complexity, removing any one degrades effectiveness, as shown in our ablation study (Section~\ref{sec:ablation}).
}

\paragraph{Target systems}
\revise[R3.O3]{%
On PostgreSQL and SQLite, \SystemName{} improves branch coverage over \ProjectName{} (Table~\ref{tab:coverage}), but fails to detect new logic bugs.
PostgreSQL and SQLite have been continuously fuzzed by tools like SQLancer (with 9.7K LOC of handwritten generators for SQLite alone) for several years, and are widely recognized as among the most thoroughly tested open-source DBMSs.
Not finding new bugs in these systems is consistent with the extensive prior testing effort.
For example, DDLCheck, QPG and TQS~\cite{song2025detecting, ba2023testing,tang2023detecting} likewise reported no new bugs on them.
}

\paragraph{Future work}
\revise[R3.O1 R3.Q1]{%
Direct LLM-based SQL generation has limitations such as hallucination, low throughput, high cost, and incompatibility with established test oracles, yet interest in self-evolving agents that could synthesize SQL generators automatically is growing.
We initially explored this direction (see Section~\ref{sec:LLM-generator}), but at the time, LLMs struggled with internal DBMS state and the complex SQL grammar required for robust generator synthesis.
With rapid progress in LLM capabilities, this direction merits renewed investigation: future agentic workflows could let an LLM synthesize, analyze, and continually refine test generators autonomously using execution feedback and external knowledge.
}

\section{Related Work}\label{sec:relatedwork}

\paragraph{LLM-aided testing}
Recent work has leveraged LLMs for automated software testing across multiple domains.
Fuzz4All~\cite{xia2024fuzz4all} is a universal fuzzing framework that uses LLM-driven prompt generation and mutation to generate diverse test inputs in various programming languages.
WhiteFox~\cite{yang2024whitefox} is an LLM-based white-box compiler fuzzer where one model analyzes compiler optimization passes to derive input requirements and another model generates test programs.
QTRAN~\cite{lin2025qtran} applies LLMs to translate query pairs across SQL dialects for metamorphic testing, fine-tuning a model to preserve oracle relations during mutation.
These methods use LLMs to generate inputs directly, resulting in low throughput, and LLM hallucinations can lead to false bug reports, as reported in the QTRAN work.
Conversely, \SystemName{} separates learning and testing, and generates inputs through traditional generators, making the process efficient and cost-effective.
KernelGPT~\cite{yang2025kernelgpt} is an approach to enhance OS kernel fuzzing by leveraging LLMs to synthesize system call specifications for fuzzers.
MetaMut~\cite{ou2025mutators} is an LLM-guided mutation testing approach that integrates compiler domain knowledge into prompts to automatically generate high-quality mutators for compiler fuzzing.
As in \SystemName{}, MetaMut invokes the LLM only during an initial phase and does not use it as a direct test case generator.

\paragraph{Test case generation for DBMSs}
Existing methods for generating test cases for DBMSs can be divided into two categories: mutation-based testing and generator-based testing.
Mutation-based testing for DBMSs is based on SQL-specific mutators to mutate existing SQL statements.
Griffin~\cite{fu2023griffin} uses a grammar-free SQL fuzzing approach that replaces hand-written grammars by summarizing the database state in a lightweight metadata graph to guide semantically correct query mutations.
BuzzBee~\cite{yang2024towards} extends DBMS fuzzing to cover multiple database models.
Sedar~\cite{fu2024sedar} improves the effectiveness of mutation-based testing by transferring SQL seeds across DBMS dialects, thus obtaining high-quality diversified inputs.
None of the above mutation-based testing approaches can be applied to find logic bugs due to semantic constraints from test oracles.
In contrast, generator-based approaches construct queries based on pre-defined rules.
SQLsmith~\cite{sqlsmith} is a representative tool, which utilizes schema metadata to generate well-formed random SQL queries.
SQLancer~\cite{rigger2020testing, rigger2020detecting, rigger2020finding} also generates queries based on various hand-written SQL generators. 
Both tools are built on manually implemented generators, and \SystemName{} can automatically augment these generators without human supervision.

\paragraph{DBMS test oracles}
\SystemName{} can use existing test oracles.
In this work, we implemented TLP and NoREC.
TLP~\cite{rigger2020finding} detects logic bugs by executing a query and checking that its result matches the combined results of three queries that split the rows into separate parts.
NoREC~\cite{rigger2020detecting} detects logic bugs by executing a query that is receptive to optimizations and comparing its result to an equivalent version that is unlikely to be optimized.
Various other test oracles could be integrated by \SystemName{}.
EET~\cite{jiang2024detecting} rewrites each query through equivalent transformations and checks that the rewritten query returns the same results as the original one.
CODDTest~\cite{zhang2025constant} leverages compiler optimizations to find logic bugs in DBMSs, especially in advanced features like subqueries.
APOLLO~\cite{jung2019apollo} finds performance regressions by running random queries on different versions of the same DBMS, and flagging cases where the newer version runs much slower.
Several other test oracles cannot be supported in a dialect-agnostic manner as they require DBMS-specific information. For example, CERT~\cite{ba2023finding} requires manual effort to implement query plan parsers, with format usually differing across DBMSs, to find unexpected differences in the cardinality estimator.
Radar~\cite{song2024detecting} requires manual effort in designing and analyzing the metadata constraints for the raw database generation.
Conceptually, \SystemName{} could be applied to augment the generators of all these oracles, thereby improving their bug-detection capabilities.

\paragraph{LLM- and AI-empowered systems}
LLMs have recently been leveraged in database management for tasks such as query rewriting, SQL dialect translation, and text-to-SQL conversion.
DB-GPT~\cite{zhou2024db} treats the LLM as the ``brain'' of the DBMS that can adaptively handle tasks like automatic query reformulation and index recommendation.
LLM-R\textsuperscript{2}~\cite{li2024llmr2} is a rule-based rewrite framework that uses an LLM to recommend rewrite rules drawn from existing SQL rewriting platforms.
CrackSQL~\cite{zhou2025cracksql} combines rule-based techniques with LLMs to translate between SQL dialects.
FinSQL~\cite{zhang2024finsql} is a model-agnostic, LLM-based text-to-SQL framework for financial data analysis.
Similar to the above systems, \SystemName{} also leverages LLM knowledge in database management, but focuses on enhancing DBMS testing tools.

CatSQL~\cite{fu2023catsql} combines rule-based SQL sketch templates with deep learning models to fill in query details, improving the accuracy and reliability of NL2SQL translation.
\SystemName{} shares a similar idea with CatSQL in that both leverage a language model to fill SQL sketches; however, \SystemName{} differs in the following respects.
First, CatSQL focuses on \texttt{SELECT} queries, whereas \SystemName{} supports all kinds of SQL statements.
Second, in CatSQL, the filled sketches are final system outputs. In contrast, \SystemName{} uses the filled sketches to improve its generator, and the generator generates outputs.
Third, CatSQL translates natural language queries to SQL queries, while \SystemName{} aims to learn features of different dialects for DBMS testing.

\revise[R1.M R2.O1]{%
\paragraph{Validation of LLM-generated SQL}
Existing works leverage different approaches to validate LLM-generated SQL.
LITHE~\cite{dharwada2025lithe} validates semantic equivalence between LLM-suggested rewrites and original queries.
QO-Verify~\cite{narasayya2026leveraging} uses query optimizers to verify LLM-based rewrites.
SQLDriller~\cite{yang2025automated} generates counterexample database instances to check that a natural-language query and its candidate SQL translation agree on execution.
Most closely related to \SystemName{} is the concurrent work Argus~\cite{mang2025automated}, which prompts an LLM to synthesize equivalent query pairs as test oracles and validates the equivalence using a SQL equivalence prover.
Argus addresses the long-standing challenge of designing high-quality test oracles for automated DBMS testing.
In contrast, \SystemName{} uses the LLM to synthesize test inputs fed into a traditional generator and paired with existing oracles such as TLP and NoREC, targeting the problem of improving test input diversity across SQL dialects, and supports a broader range of SQL features (\emph{e.g.}, DML statements) than Argus.
The two directions are complementary: oracles generated by Argus could be combined with \SystemName{}'s generator augmentation to further enhance testing effectiveness.
}

\section{Conclusion}
Existing SQL generators for logic bug detection require significant human effort for adaptation to various DBMSs.
In this paper, we have presented an LLM-aided approach, \SystemName{}, for automatically augmenting these existing generators.
To the best of our knowledge, this is the first approach that utilizes LLMs within an automated testing tool for DBMSs.
Besides, we introduce the notion of SQL sketches to incorporate SQL features automatically into query generators.
Conceptually, we believe that techniques inspired by our approach could be applied to efficiently utilize LLMs in the context of migration of queries to other dialects, text-to-SQL, or autocompletion in SQL editors.
\SystemName{} has discovered \OverallBugs{} previously unknown bugs in five DBMSs, \FixedBugs{} of which have been fixed.
While manually written dialect-specific generators still outperform \SystemName{} in some cases---3\% more branch coverage on SQLite---the difference is small.
Furthermore, our approach can be applied to any SQL-based DBMS without any implementation effort.

\begin{acks}
We would like to sincerely thank all the anonymous reviewers for their valuable feedback and insights that helped us improve the quality of this paper. We want to thank all the DBMS developers for responding to our bug reports and for analyzing and fixing the issues we identified. This research is supported by the Google South \& Southeast Asia Research Award, the National Research Foundation, Singapore and the Cyber Security Agency of Singapore under its National Cybersecurity R\&D Programme (Fuzz Testing), and a Singapore Ministry of Education (MOE) Academic Research Fund (AcRF) Tier 1 grant. Any opinions, findings and conclusions, or recommendations expressed in this material are those of the author(s) and do not reflect the views of Google, the National Research Foundation, Singapore, the Cyber Security Agency of Singapore, or the Singapore Ministry of Education.
\end{acks}

\bibliographystyle{ACM-Reference-Format}
\bibliography{sample-base}
\end{document}